\documentclass[preprint,nofootinbib,showkeys]{revtex4-1}
\usepackage{amsmath}
\usepackage{amssymb}
\usepackage{graphicx}
\makeatletter
\usepackage{hyperref}
\hypersetup{
	colorlinks = true,
	linkcolor = {blue},
	urlcolor = {black},
	citecolor = {blue}
}

\makeatother

\begin{document}
\count\footins = 1000
\title{Analysis on the black hole formations inside old neutron stars by
isospin-violating dark matter with self-interaction}
\author{Guey-Lin Lin$^1$}
\email{glin@cc.nctu.edu.tw}
\author{Yen-Hsun Lin$^2$}
\email{yenhsun@gate.sinica.edu.tw}
\affiliation{$^1$Institute of Physics, National Chiao Tung University, Hsinchu 300, Taiwan \\
	$^2$Institute of Physics, Academia Sinica, Taipei 115, Taiwan}

\begin{abstract}
Fermionic dark matter (DM) with attractive self-interaction is possible
to form black holes (BH) inside the Gyr-old neutron stars (NS). Therefore
by observing such NS corresponding to their adjacent DM environments
can place bounds on DM properties, eg.~DM-baryon cross section $\sigma_{\chi b}$,
DM mass $m_{\chi}$, dark coupling $\alpha_\chi$ and mediator mass $m_\phi$.
In case of isospin violation, DM couples to neutron and proton in different strengths.
Even NS is composed of protons roughly one to two percent of the total baryons, the contribution
from protons to the DM capture rate  could be drastically
changed in the presence of isospin violation. We demonstrate that this effect can be important in certain cases.
On the other hand,
DM-forming BH inside the star is subject to many criteria and the
underlying dynamics is rich with interesting features. We also systematically
review the relevant physics based on the virial equation. Moreover,
an accompanied python package \texttt{dm2nsbh} to realize the mechanism
is also released on the github for other relevant research.
\end{abstract}
\maketitle

\section{Introduction}

Dark matter (DM) composites one-fifth of the Universe but its particle
essence remains elusive. 
To 
discern the nature of DM is a
great challenge in modern physics. Plethora of experiments to detect
the signal from dark sector either from the direct interaction between
DM and the Standard Model (SM) particles \cite{Aad:2015zva,Abdallah:2015ter,Aalbers:2016jon,Akerib:2016vxi,Amole:2017dex,Akerib:2017kat,Aprile:2017iyp,Aprile:2018dbl} or from the indirect
measurement of the events produced by DM annihilation or decay \cite{Aartsen:2014oha,Choi:2015ara,Aartsen:2016zhm,Aguilar:2015ctt,TheFermi-LAT:2017vmf,Ambrosi:2017wek}
are undergoing.

An interesting feature is that DM particles can accrete in centers
of stellar objects through the energy loss due to DM-baryon interaction
characterizing by the scattering cross section $\sigma_{\chi b}$
where $b\equiv n,p$, the neutron and proton, respectively. Additionally,
if the captured DMs do annihilate into SM particles, they could provide
viable signals to be detected by the terrestrial detectors. The case for
the solar-capture DM has been studied recently in Refs.~\cite{Chen:2014oaa,Kong:2014mia,Chen:2015bwa,Chen:2015uha,Chen:2015poa,Catena:2016ckl,Garani:2017jcj,Fornengo:2017lax,Chen:2018lsk,Gaidau:2018yws}.
On the other hand, if captures happen in the compact stars such
as neutron stars (NS), the annihilation products could potentially
cause the surface temperature of the star deviating from the standard prediction
\cite{Kouvaris:2007ay,deLavallaz:2010wp,Kouvaris:2010vv,Kouvaris:2010jy,Leung:2011zz,McDermott:2011jp,Kouvaris:2011gb,Guver:2012ba,Bramante:2013nma,Tolos:2015qra,Bramante:2017xlb,Baryakhtar:2017dbj,Raj:2017wrv,Ellis:2017jgp,Ellis:2018bkr,Bell:2018pkk,Garani:2018kkd,Hamaguchi:2019oev,Dasgupta:2019juq,Acevedo:2019agu,Joglekar:2019vzy,Chen:2018ohx}. This 
can be used to constrain the DM properties. 

A different scenario is the asymmetric DM (ADM) 
as reviewed in Refs.~\cite{Kaplan:2009ag,Petraki:2013wwa,Zurek:2013wia}.
Unlike the
previous case, there is no anti-DM left in the current Universe 
to enable the annihilation. Without the depletion due to annihilation,
the number of captured DM particles grow unlimited until collapsing
into a black hole (BH) and consume the entire star. Investigations
on this issue by including the BH evaporation as well as 
BH formation delay due to star rotation are done recently \cite{Kouvaris:2013kra}. Bounds on the DM mass $m_{\chi}$ versus $\sigma_{\chi b}$
are 
derived for different types of DM, eg.~fermionic or bosonic and
with or without Bose-Einstein condensation (BEC) \cite{Kouvaris:2010vv,McDermott:2011jp,Colpi:1986ye,Kouvaris:2015rea,Boehmer:2007um,Eby:2015hsq,Zheng:2014fya}.

Besides DM-baryon interaction, it is suggested that DM could be self-interacting.
The self-interacting DM (SIDM) is introduced to alleviate the discrepancies between
$N$-body simulations and astrophysical observations at the small scale, i.e., the core-cusp problem, missing satellite
problem, too-big-to-fail problem and diversity problem of galactic rotation curves. See Refs.~\cite{Bullock:2017xww,Tulin:2017ara} for
comprehensive reviews and the references therein.
The conditions given in Refs.~\cite{Randall:2007ph,Walker:2011zu,BoylanKolchin:2011de,BoylanKolchin:2011dk,Elbert:2014bma}, such as  
\begin{equation}
0.1~{\rm cm}^2~{\rm g}^{-1}<\sigma_{\chi \chi}/m_\chi<10~{\rm cm}^2~{\rm g}^{-1},\label{eq:sidm_constr}
\end{equation}
can mitigate these small scale problems where $\sigma_{\chi\chi}$ is the DM self-interaction cross section.
In principle, SIDM can be either attractive or repulsive. In this work, we only focus on
the attractive SIDM. The repulsive interaction is not considered here since it would counterbalance the gravitational contraction
in addition to the Fermi pressure. These two effects together shall forbid DM to collapse into a BH inside NS.

Therefore, in the following content, we analyze the scenario with fermionic ADM captured by NS.
The captured DM particles could form a BH
and consume the entire star in the presence of attractive self-interaction. 
Although our analysis is based on a model-independent perspective,
we shall provide a feasible phenomenological 
scenario to justify our analysis in the following sections.
Thus, observations 
of very old NS in a DM-rich 
environment can place bound on the strengths of DM self- as well as
DM-baryon interactions.
A systematic algorithm to
determine if all the criteria for BH formation are met is provided
in Ref.~\cite{Bramante:2013nma}. We 
adopt such an algorithm in this
study, and briefly summarize the technical details as well as the
relevant physical meanings in the following sections. For completeness,
an accompanied python package \texttt{dm2nsbh} 
which implements the method
in Ref.~\cite{Bramante:2013nma} is also released on the github \cite{yhl_git}.

In the following content we always use $\hbar=c=k_{B}=1$ and the
paper is organized as follows. 
In Sec.~\ref{sec:model}, a simple phenomenological framework is presented to
justify our analysis.
In Sec.~\ref{sec:DM_acceretion}, we provide the
general formalism for DM accretion in the NS.
In Sec.~\ref{sec:BH_formation},
the criteria for DM to form a BH in the NS are discussed. In Sec.~\ref{sec:DM_isospin},
NS sensitivities to $\sigma_{\chi b}$ are derived. We summarize in Sec.~\ref{sec:Summary}.

\section{Phenomenological framework\label{sec:model}}

\begin{figure}
	\begin{centering}
		\begin{tabular}{ccc}
			\includegraphics[width=0.25\textwidth]{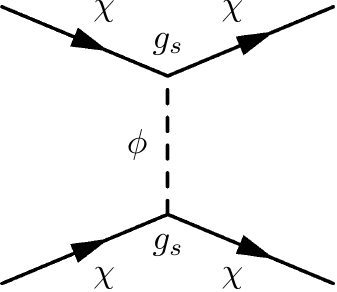} &  &
			\includegraphics[width=0.25\textwidth]{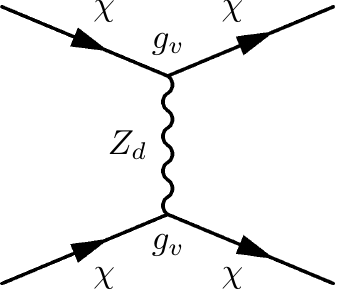}\tabularnewline
			(a)&&(b) \tabularnewline
		\end{tabular}
		
		\par\end{centering}
	\caption{\label{fig:feyn}Feynman diagrams for DM self-interaction, mediators are $\phi$ for
	 (a) and $Z_d$ for (b).}
\end{figure}

Let us assume that the $U(1)_d$ gauge boson $Z_d^{\mu}$ gets the mass from its coupling to complex scalar field $\Phi$.	
After spontaneous symmetry breaking (SSB), we have $\Phi=(\phi +v_d+i\sigma)/\sqrt{2}$, where $\phi$ and $v_d$
are dark Higgs field and its vacuum expectation value, respectively, $\sigma$ is the Goldstone boson field. The dark Higgs $\phi$ and the $U(1)_d$ gauge boson ${Z_d}$
can acquire masses through the above symmetry breaking such that $m_\phi^2 =2\lambda v_d^2$ and $m_{Z_d}^2= g_d^2v_d^2$ with $\lambda$ the dark Higgs self-coupling constant in the scalar potential $V(\Phi)=\mu^2(\Phi^*\Phi)+\lambda (\Phi^*\Phi)^2$ and $g_d$ the dark charge carried by dark Higgs.
DM field $\chi$, taken as Dirac fermion, is the linear combination of fermionic fields $\xi$ and $\eta$ in the hidden sector with the former carrying $U(1)_d$ dark charge $g_d$ and the latter a $U(1)_d$ singlet.  The Lagrangian for gauge and Yukawa interactions can be written as  $\mathcal{L}_{\rm int}=\mathcal{L}_G+\mathcal{L}_Y$ where
\begin{equation}
\mathcal{L}_G=i\bar{f}\gamma_{\mu}D^{\mu}f+(D_{\mu}\Phi^*)(D^{\mu}\Phi),
\end{equation}
with $f=(\xi, \eta)_T$, $D^{\mu}(\xi,\Phi)=(\partial^{\mu}-ig_dZ_d^{\mu})(\xi,\Phi)$, and $D^{\mu}\eta=\partial^{\mu}\eta$, while
\begin{equation}
\mathcal{L}_Y=-m_\xi \bar{\xi} \xi - m_\eta \bar{\eta}\eta - y\Phi^*\bar{\xi} \eta - y\Phi \bar{\eta} \xi,
\end{equation}
with $y$ the Yukawa coupling constant. The mass terms $-m_\xi \bar{\xi} \xi - m_\eta \bar{\eta}\eta$ are included in 
$\mathcal{L}_Y$ so that this Lagrangian contains complete mass terms after SSB. The details of the diagonalization  are discussed in~Appendix \ref{sec:model_detail}.
In terms of mass eigenstates, DM self-interaction can be generated from scalar and vector interactions of DM: 
\begin{equation}
\mathcal{L}_S=g_s \phi \bar{\chi}\chi\quad{\rm and}\quad\mathcal{L}_V =g_{v}\bar{\chi}\gamma_{\mu}\chi Z_d^\mu, 
\end{equation}
where $g_s=-y\sin 2\theta/\sqrt{2}$ and $g_v=g_d \sin^2\theta$ with $\theta$ 
the rotation angle between gauge and mass bases.
The corresponding Feynman diagrams are depicted in Fig.~\ref{fig:feyn}.

\subsection{DM self-interaction}

The amplitudes for DM self-interaction shown in Fig.~\ref{fig:feyn} are given by
\begin{equation}
\mathcal{M}_a \sim \frac{g_s^2}{m_\phi^2}=\frac{g_s^2}{2\lambda v_d^2}\quad{\rm and}\quad
\mathcal{M}_b \sim \frac{g_v^2}{m_{Z_d}^2}=\frac{\sin^4\theta}{ v_d^2}.
\label{eq:scalar_dominance}
\end{equation}
in the zero-momentum transfer limit. 
The potentials induced by diagram (a) and (b) are attractive and repulsive, respectively.
Our interested scenario corresponds to $\mathcal{M}_a\gg \mathcal{M}_b$, i.e., $g_s^2\gg 2\lambda \sin^4\theta$, or equivalently, $y^2\gg \lambda\tan^2\theta$.
In this case, DM self-interaction can be described by the Yukawa potential
\begin{equation}
V(r)=-\frac{\alpha_{\chi}}{r}e^{-m_{\phi}r},\label{eq:Yukawa}
\end{equation}
with
\begin{equation}
\alpha_{\chi}\equiv \frac{g_s^2}{4\pi}\label{eq:ax}
\end{equation}
the dark fine structure constant. 
In the non-relativistic limit, DM self-interaction
cross section $\sigma_{\chi\chi}$ can be obtained through solving the Schr\"{o}dinger's
equation with the potential $V(r)$.
The theoretical frameworks \cite{Buckley:2009in,Tulin:2013teo,Wise:2014jva} and astrophysical
constraints on $\sigma_{\chi\chi}$ \cite{Kamada:2016euw,Robertson:2017mgj,Oman:2015xda,Elbert:2016dbb}
have been 
extensively investigated.

The connection between Eqs.~(\ref{eq:sidm_constr}) and (\ref{eq:Yukawa}) is thus established. Following Ref.~\cite{Tulin:2013teo}, in the Born limit, $\alpha_\chi m_\chi/m_\phi \ll 1$, the self-interaction cross section is given by
\begin{equation}
\sigma_{\chi\chi}^{\rm Born}=\frac{8\pi \alpha_\chi^2}{m_\chi^2 v^4}\left[\ln\left(1+\frac{m_\chi^2 v^2}{m_\phi^2}\right)-\frac{m_\chi^2 v^2}{m_\chi^2 v^2+m_\phi^2} \right],\label{eq:sigxx_born}
\end{equation}
where $v$ is the DM relative velocity during the collision.
Beyond the Born region, we have $\sigma_{\chi\chi}$ in the case of classical limit, $m_\chi v/m_\phi \gg 1$, with attractive Yukawa potential,
\begin{equation}
\sigma_{\chi\chi}^{{\rm class}}=\begin{cases}
\frac{4\pi}{m_{\phi}^{2}}\beta^{2}\ln(1+\beta^{-1}), & \beta\lesssim0.1\\
\frac{8\pi}{m_{\phi}^{2}}\beta^{2}/(1+1.5\beta^{1.65}), & 0.1\lesssim\beta\lesssim10^{3}\\
\frac{\pi}{m_{\phi}^{2}}(\ln\beta+1-\frac{1}{2}\ln^{-1}\beta)^{2}, & \beta\gtrsim10^{3},
\end{cases}\label{eq:sigxx_class}
\end{equation}
where $\beta=2\alpha_\chi m_\phi/(m_\chi v^2)$. 
For $m_\chi/m_\phi<1$, we have an approximated expression for $\sigma_{\chi\chi}$ in the non-perturbative region,
\begin{equation}
\sigma_{\chi \chi}^{\rm non-peturb}=\frac{16\pi}{m_\chi^2 v^2}\sin^2\delta_0,\label{eq:sigxx_hulthen}
\end{equation}
where the exact form of phase shift $\delta_0$ is given by Eq.~(A5) in Ref.~\cite{Tulin:2013teo}.
The derivations of Eqs.~(\ref{eq:sigxx_born}-\ref{eq:sigxx_hulthen}) are beyond the scope of the paper.
We refer the reader to Ref.~\cite{Tulin:2013teo} and the references therein for mathematical details.

\subsection{DM-baryon interaction}
The portal that bridges the dark sector and SM can be established by $Z_d$ mixing with
SM photon and $Z$ boson
through kinetic and mass mixing terms\footnote{The $Z-Z_d$ mass mixing can introduce corrections to both $Z$ and $Z_d$ masses. On the other hand, if the mixing is of the same order as $g_d^2v_d^2$, the correction to the $Z_d$ mass is negligible and the arguments following Eq.~(\ref{eq:scalar_dominance}) remain valid.}, 
respectively \cite{Davoudiasl:2012ag}.
These mixings provide the interactions 
\begin{equation}
\mathcal{L}_{\rm mix} = \left(\varepsilon_\gamma e J_\mu^{\rm EM}+\varepsilon_Z \frac{g_2}{c_W}J_\mu^{\rm NC}\right)Z_d^\mu, \label{eq:L_int}
\end{equation}
where $e$, $\varepsilon_\gamma$ and $\varepsilon_Z$ are SM electric charge, kinetic mixing and mass mixing parameters, respectively, $g_2$ is the $SU(2)_L$ coupling, $J_\mu^{\rm EM,NC}$ are the SM electromagnetic and weak neutral currents, respectively, and $c_W\equiv \cos \theta_W$ with $\theta_W$ the Weinberg angle. 
\begin{figure}
	\begin{centering}
		\includegraphics[width=0.25\textwidth]{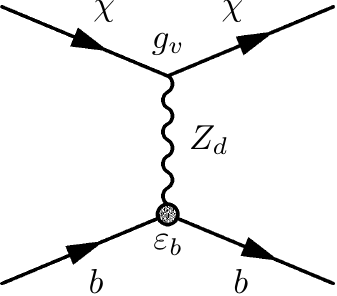}
		\par\end{centering}
	\caption{\label{fig:DM-b}Feynman diagrams for DM-baryon interaction where $b$ stands for
		either neutron $n$ or proton $p$.}
\end{figure}
Furthermore, from Eq.~(\ref{eq:L_int}), the DM-baryon interaction 
is given by
\begin{equation}
\mathcal{L}_{\rm int}=\sum_{b=n,p}e\varepsilon_b \bar{b}\gamma_\mu  Z_d^\mu b
\end{equation}
where $\varepsilon_{n,p}$ are the couplings to neutron and proton respectively. 
The Feynman diagram is shown in Fig.~\ref{fig:DM-b}.
The couplings $\varepsilon_{n,p}$ are related to $\varepsilon_{\gamma}$ and  $\varepsilon_{Z}$ by \cite{Kaplinghat:2013yxa}
\begin{align}
\varepsilon_{n} & =-\frac{\varepsilon_{Z}}{4s_{W}c_{W}}\approx-0.6\varepsilon_{Z},\label{eq:en}\\
\varepsilon_{p} & =\varepsilon_{\gamma}+\frac{\varepsilon_{Z}}{4s_{W}c_{W}}(1-4s_{W}^{2})\approx\varepsilon_{\gamma}+0.05\varepsilon_{Z}.\label{eq:ep}
\end{align}
The corresponding DM-baryon cross section $\sigma_{\chi n,p}$ is proportional to $\varepsilon_{n,p}^2$.
We 
note that $\varepsilon_n$ and $\varepsilon_p$ are generally not identical, i.e., isospin symmetry is violated. 
We point out that $\phi$ can also mix with SM Higgs via scalar mixing $\varepsilon_h$. 
Since our interested $m_\phi$ range is around sub-GeV, $\phi$ can
decay into SM particles via such a portal.
BBN requires $\varepsilon_h\gtrsim 10^{-5}$. However, this parameter range has already been excluded by direct searches \cite{Kaplinghat:2013yxa}.
Thus, we ignore such a mixing in this paper.

The above phenomenological discussions provide a foundation for
self-attracting and isospin violating DM.
We will proceed our analysis later in a model-independent way.

\section{Accretion of dark matter onto a neutron star\label{sec:DM_acceretion}}

\subsection{General formalism for DM evolution}

When the halo DM particles scatter with NS and lose significant
amount of energies, they are gravitationally trapped in the star.
The evolution of DM number $N_{\chi}$ in NS can be characterized
by the differential equation
\begin{equation}
\frac{dN_{\chi}(t)}{dt}=C_{c}-C_{e}N_{\chi}(t),\label{eq:evo_eq}
\end{equation}
where $C_{c}$ is the NS capture rate due to DM-baryon scattering
and $C_{e}$ is the evaporation rate. It is argued that $C_{e}$ is negligible
unless $m_{\chi}\lesssim\mathcal{O}({\rm keV})$ \cite{Garani:2018kkd}.
In our interested parameter space, 
$C_{e}$ can be ignored so that $N_{\chi}(t)$ is given by 
\begin{equation}
N_{\chi}(t)=C_{c}t.\label{eq:Nx}
\end{equation}
Thus, DM would accumulate without limit in the star. As a remark, it
was also pointed out in Refs.~\cite{Guver:2012ba,Chen:2018ohx} that DM self-capture rate is
generally negligible unless $\sigma_{\chi b}\lesssim10^{-55}\,{\rm cm}^{2}$.
Thus it is reasonable to neglect the DM self-capture in this study.

\subsection{DM capture rate of NS}

Initially the DM capture rate $C_{c}$ for NS was inferred from 
earlier studies with respect to DM captures in the Sun and Earth 
\cite{Gould:1987ww,Gould:1987ir,Gould:1987ju,Busoni:2013kaa,Garani:2017jcj}. In recent studies \cite{McDermott:2011jp,Bell:2013xk},
the Pauli blocking effect in the DM-baryon scattering was partially
considered and the NS was assumed having a constant density. In this
work,we use the numerical data provided by the authors of Ref.~\cite{priv} for calculating $C_c$. For completeness,
we briefly summarize the method here.

The expression for NS capture rate in the degenerate medium is
given by
\begin{equation}
C_{c}=\int_{0}^{R_{0}}4\pi r^{2}dr\int_{0}^{\infty}\left(\frac{\rho_{\chi}}{m_{\chi}}\right)\frac{f(u)}{u}w(r)du\int_{0}^{v_{{\rm esc}}(r)}R^{-}(w\to v)dv\label{eq:Cc}
\end{equation}
where $\rho_{\chi}$ is the local DM density near the NS, $f(u)$
the DM velocity distribution in the NS rest frame and assumed to be
Maxwell-Boltzmannian, $v_{{\rm esc}}(r)$ the escape velocity at layer
$r$ of the NS and $w=\sqrt{u^{2}+v_{{\rm esc}}^{2}(r)}$ the velocity
of DM falling into layer $r$. The quantity $R^{-}(w\to v)$ is the
DM differential scattering rate 
from the initial velocity $w$ to a smaller final velocity $v$. It is explicitly written
as\footnote{Assuming $\sigma_{\chi b}$ is 
velocity-independent, we have the factorization  $d\sigma_{\chi b}/dv=\sigma_{\chi b}K(v)$ where $K(v)$ is certain function of $v$.
The exact form of $K(v)$ is not relevant to our discussions. For readers who are interested in evaluating Eqs.~(\ref{eq:Cc},\ref{eq:R-}) numerically, we refer them to the appendix of Ref.~\cite{Garani:2018kkd}.
}
\begin{equation}
R^{-}(w\to v)=\sum_{b}\int n_b(r)\frac{d\sigma_{\chi b}}{dv}|w-u_b|f_{b}(E_{b},r)[1-f_{b}(E_{b}+q_{0},r)]d^{3}u_b\label{eq:R-}
\end{equation}
where $b=n,p$ stands for neutron and proton respectively. The summation
indicates both neutrons and protons in the NS contributing to the capture
of DM and can be treated separately. The quantity $n_b(r)$ is the baryon number density,
$q_{0}=m_{\chi}(w^{2}-v^{2})/2$ the DM energy loss 
for single scattering and $u_b$ the baryon velocity.\footnote{Determined by $E_b=\sqrt{m_b^2+(\gamma m_b u_b)^2}$ where $\gamma=1/\sqrt{1-u_b^2}$ is the Lorentz factor.}
The Fermi-Dirac distribution for baryon energy is given by
\begin{equation}
f_{b}(E_{b},r)=\frac{1}{e^{(E_{b}-\mu_{F}(r))/T_{{\rm NS}}(r)}+1},
\end{equation}
with $\mu_{F}(r)$ the baryon chemical potential and $T_{{\rm NS}}(r)$ the NS temperature at layer $r$.

\begin{figure}
\begin{centering}
\includegraphics[width=0.45\textwidth]{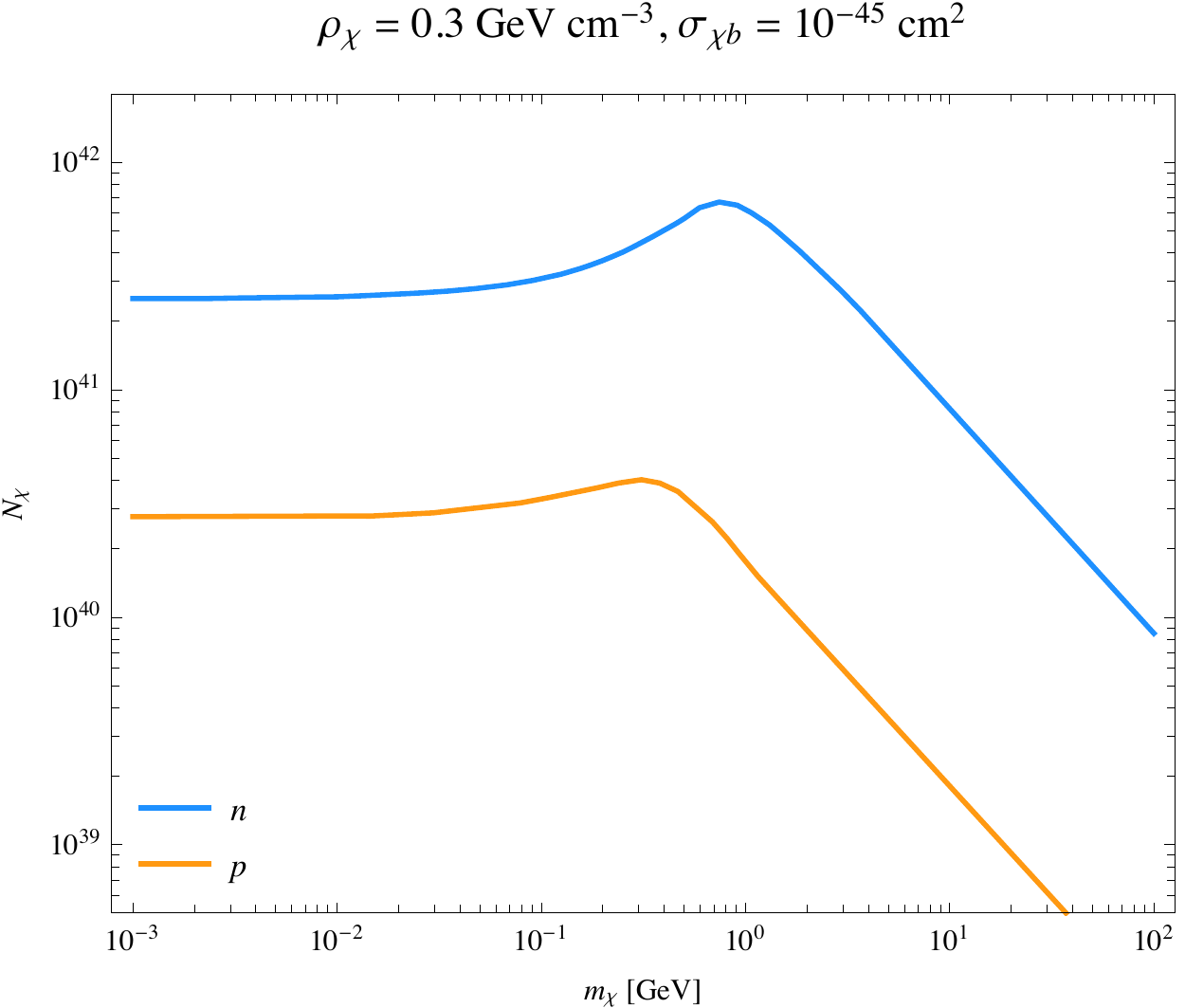}
\par\end{centering}
\caption{\label{fig:Nx}The contributions from neutrons and protons to the
captured DM number $N_{\chi}$ in an NS with age $t_{{\rm age}}=5\,{\rm Gyrs}$
and $\sigma_{\chi b}=10^{-45}\,{\rm cm}^{2}$. The environment has
$\rho_{\chi}=0.3\,{\rm GeV\,cm}^{-3}$ and the DM velocity dispersion
$\bar{v}=220\,{\rm km\,s}^{-1}$.}
\end{figure}

As a remark, the capture rates by neutrons and protons are not
identical in NS. 
In fact, protons only account for $2.7\%$ of total baryons in NS. Furthermore the chemical potential and radial density distributions of protons differ from 
those of neutrons 
\cite{Garani:2018kkd}.\footnote{Here isospin symmetry for DM-nucleon coupling is assumed. The case with isospin violation will be addressed in the next subsection. }
Fig.~\ref{fig:Nx} shows $N_{\chi}$
captured by different baryons in NS
where the flat regions are due to Pauli blocking suppression. However,
for $m_{\chi}\gtrsim1\,{\rm GeV}$, the Pauli blocking effect becomes negligible so that
the corresponding capture rate $C_{c}$ is proportional to $1/m_{\chi}$ as usual.
We also point out that the geometrical cross sections for DM-neutron scattering is
$\sigma_{\chi n}^{\rm geom}\approx 10^{-45}\,{\rm cm}^2$ and that for DM-proton is
$\sigma_{\chi p}^{\rm geom}\approx 35\sigma_{\chi n}^{\rm geom}$ \cite{Garani:2018kkd}. 

\subsection{Isospin violating DM}

In principle, the DM scattering cross section with nuclei $A$ can be
expressed in terms of $\varepsilon_{n,p}$, the couplings to neutron and proton
\cite{Jungman:1995df},
\begin{equation}
\sigma_{\chi A}=\frac{4}{\pi}\frac{\mu_{A}^{2}}{\Lambda^4}[Z\varepsilon_{p}+(A-Z)\varepsilon_{n}]^{2}\label{eq:effective_couple}
\end{equation}
where $\Lambda$ is certain energy scale, $\mu_{A}=m_{A}m_{\chi}/(m_{A}+m_{\chi})$ is the reduced mass
and $A$ and $Z$ are the mass and atomic numbers respectively.\footnote{We introduce energy scale $\Lambda$ to ensure 
Eq.~(\ref{eq:effective_couple}) carrying
the correct dimension of $[{\rm length}]^2$. From the Feynman diagram in Fig.~\ref{fig:DM-b}, one can see that $\Lambda\sim m_{Z_d}g_v^{-1/2}$.
}
Thus,
we can express $\sigma_{\chi p}$ in terms of $\sigma_{\chi n}$ and
$\varepsilon_{n,p}$ by
\begin{equation}
\sigma_{\chi p}=\left(\frac{\varepsilon_{n}}{\varepsilon_{p}}\right)^{-2}\sigma_{\chi n}.\label{eq:sigxp}
\end{equation}
Note that we have taken $m_{p}\approx m_{n}$. Since $\sigma_{\chi n,p}$ is proportional to
$\varepsilon_{n,p}^2$, the signs of these couplings are irrelevant to $C_c$. But the direct search bound
will be sensitive to these signs since DM scatters with the entire nuclei coherently, see Eq.~(\ref{eq:effective_couple}).

If $\varepsilon_{n}/\varepsilon_{p}=0.1$,
then $\sigma_{\chi p}$ is 100 times larger than $\sigma_{\chi n}$.
Even though protons only account for roughly 2.7\% of the total baryons in the
NS, their contributions to the capture rate $C_{c}$ are comparable to those of neutrons
due to the enhancement from isospin violation.

\begin{figure}
	\includegraphics[width=0.45\textwidth]{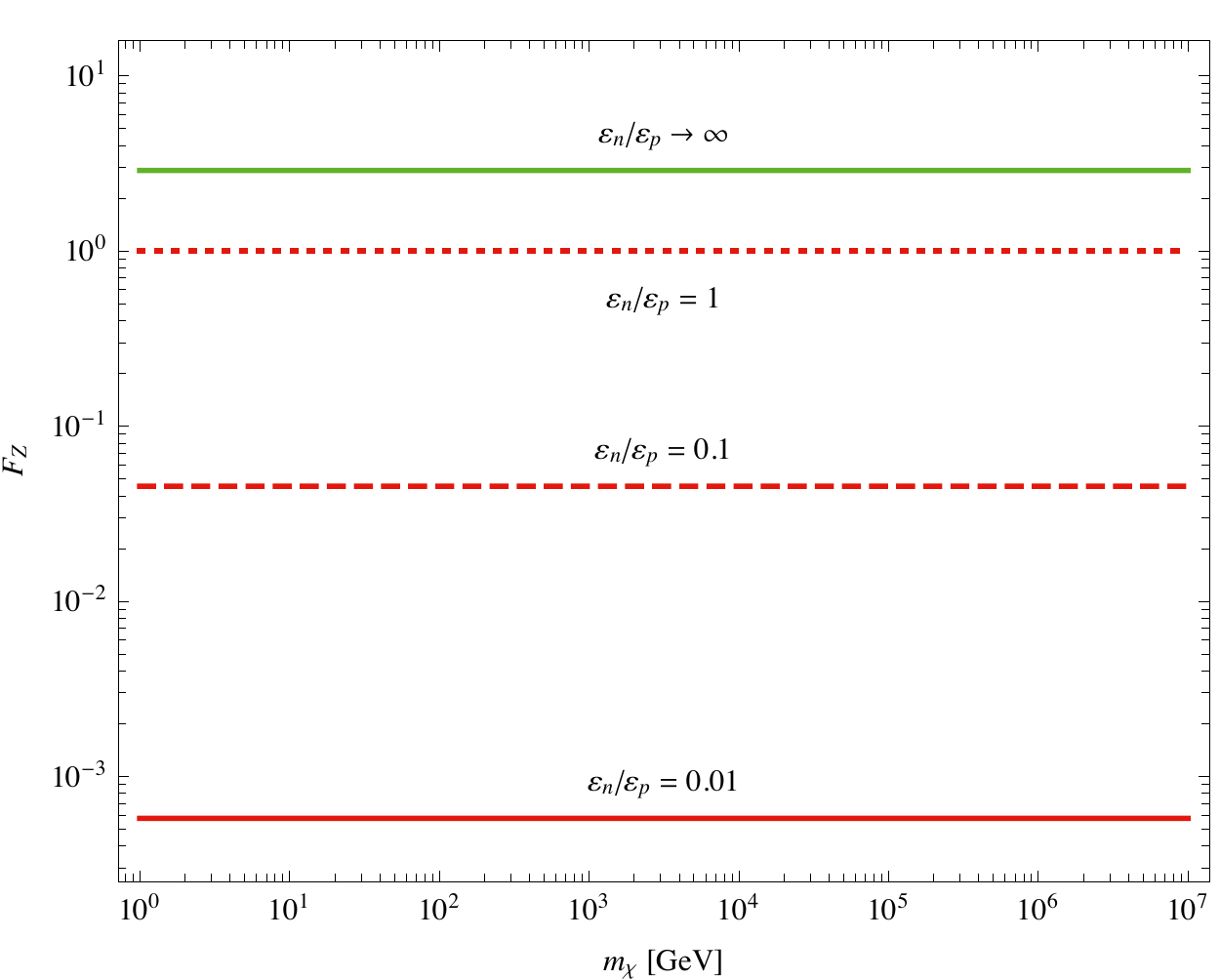}\quad
	\includegraphics[width=0.45\textwidth]{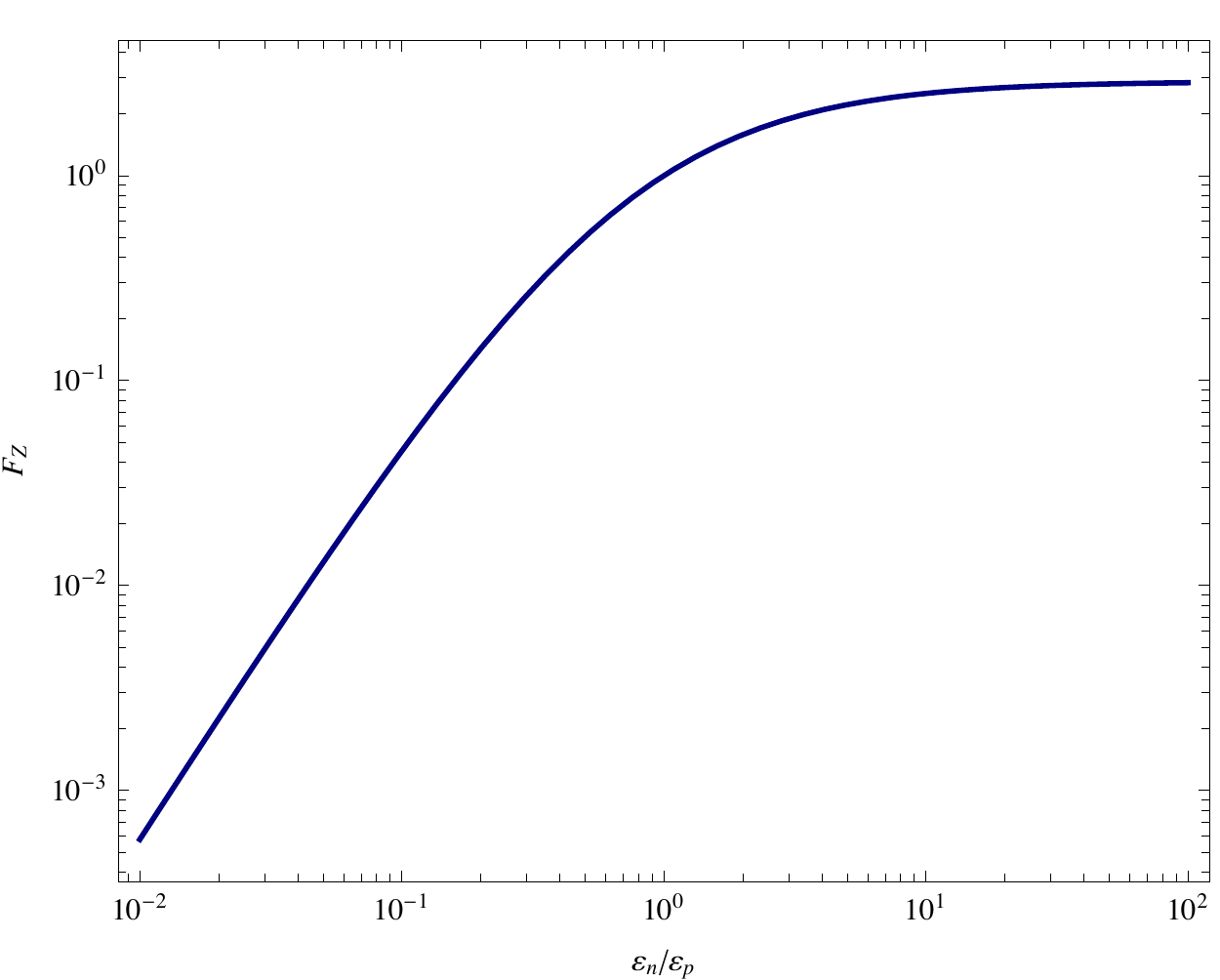}
	
	\caption{\label{fig:Fz} \textit{Left}: $F_Z$ versus $m_\chi$ with various $\varepsilon_n/\varepsilon_p$. It is clearly shown that
	$F_Z$ is independent of $m_\chi$. \textit{Right}: $F_Z$ versus $\varepsilon_n/\varepsilon_p$ with $m_\chi=100\,{\rm GeV}$.
	Different $m_\chi$ has the same results.}
\end{figure}

Without the loss of generality, we assume that those bounds on DM-baryon cross section 
obtained from direct searches are analyzed in the  isospin symmetric
case $\varepsilon_{n}/\varepsilon_{p}=1$ and are denoted as $\sigma_{\chi n}^{0}$ for
DM-neutron cross section. When isospin violation is included, the
bound on $\sigma_{\chi n}$ should be rescaled by the factor \cite{Jungman:1995df,Feng:2011vu,Lin:2014hla}
\begin{equation}
\frac{\sigma_{\chi n}}{\sigma_{\chi n}^{0}}=\frac{\sum_{i}\eta_{i}A_{i}^{2}\mu_{A}^{2}}{\sum_{i}\eta_{i}\mu_{A}^{2}(Z\varepsilon_{p}/\varepsilon_{n}+(A_{i}-Z))^{2}}\equiv F_{Z},\label{eq:Fz}
\end{equation}
where $i$ runs from all isotopes of the target element in the experiment
and $\eta_{i}$ the natural abundance of the $i$-th isotope. In this
work, we take XENON1T as the benchmark experiment and the target element
is xenon with $Z=54$. $F_{Z}$ for xenon with different $\varepsilon_{n}/\varepsilon_{p}$
are shown in Fig.~\ref{fig:Fz}. For smaller $\varepsilon_n/\varepsilon_p$, $\sigma_{\chi n}$ bound
from direct searches will be strengthen by $F_Z$. On the contrary, $\sigma_{\chi p}$ bound
will be weaken accordingly. It is also clear that $F_z$ is independent of $m_\chi$
because $\mu_A\to m_A$ in the limit $m_\chi\gg m_A$. For $m_\chi\ll m_A$, $\mu_A\to m_\chi$ so that
the $m_\chi$ dependencies cancel between the numerator and denominator.
\section{Black hole formation in neutron star\label{sec:BH_formation}}

In this section, we summarize the criteria and results for DM
collapsing into a BH in the NS from Refs.~\cite{Kouvaris:2011gb,Bramante:2013nma}.
We shall not re-derive the formulae since they are not the focus of this work.
Instead we refer the readers to original papers for technical details.
In addition, a python package \texttt{dm2nsbh} \cite{yhl_git} is provided 
for implementing the method in this section. 

When DM particles are captured, they will thermalize with the surrounding
nucleons within a short time interval and form a dark spherical cloud.
The thermal radius of the cloud $r$ can be determined by the virial
equation \cite{Bramante:2013nma}
\begin{equation}
2\langle E_{k}\rangle
=\frac{4}{3}\pi G\rho_{b}m_{\chi}r^{2}+\frac{GN_{\chi}m_{\chi}^{2}}{r}+\sum_{j}^{N_{\chi}-1}\left(\frac{\alpha_{\chi}}{r_{j}}e^{-m_{\phi}r_{j}}+\alpha_{\chi}m_{\phi}e^{-m_{\phi}r_{j}}\right).\label{eq:virial_eq}
\end{equation}
The first two terms on the RHS arise from gravitational potentials of NS
and DM, respectively, and the last term is from the Yukawa interactions between DM particles.
The parameters $E_{k}$ is the kinetic energy per DM particle, $\rho_{b}$
the core baryon density, $\alpha_{\chi}$ the dark fine structure constant defined in Eq.~(\ref{eq:ax}),
$m_{\phi}$ the mass of the scalar mediator, and $r_{j}$ the interparticle distance. 

As the number of DM $N_{\chi}$ increases with time, the last two terms on the RHS of Eq.~(\ref{eq:virial_eq}) increases with time as well. 
Once $N_{\chi}$ crosses the critical value
$N_{{\rm crit}}$ that those two terms becomes more dominant than the first term, the dark spherical cloud
supported by the kinetic energy $E_{k}$ can no longer sustain the
potentials on the RHS of Eq.~(\ref{eq:virial_eq}). Therefore,
DM particles will initiate self-gravitating and collapse. At the
moment of collapse with $N_{\chi}\gtrsim10^{36}$,
the collapse starts
from a degenerate state with the properties \cite{Bramante:2013nma}
\begin{equation}
E_{k,{\rm deg}}=\frac{(9\pi N_{\chi}/4)^{2/3}}{2m_{\chi}r_{{\rm th,deg}}^{2}}\quad{\rm and}\quad r_{{\rm th,deg}}\approx2.4\times10^{-4}\,{\rm cm}\,N_{\chi}^{1/6}\left(\frac{m_{\chi}}{{\rm GeV}}\right)^{-1/2}.
\end{equation}
On the contrary, if $N_{\chi}\lesssim10^{36}$, the collapse happens
in a non-degenerate state with 
\begin{equation}
E_{k,{\rm nondeg}}=\frac{3}{2}T_{\chi}\quad{\rm and}\quad r_{{\rm th,nondeg}}\approx250\,{\rm cm}\,\left(\frac{m_{\chi}}{{\rm GeV}}\right)^{-1/2}
\end{equation}
where $T_{\chi}$ is the DM temperature. 
It is reasonable to take $T_{\chi}=T_{{\rm NS}}=10^{5}\,{\rm K}$

\subsection{State after collapse\label{subsec:after_coll}}

When DM initiates self-gravitating, there are several phases depending
on $\alpha_{\chi}$ and $m_{\phi}/m_{\chi}$. We briefly summarize
as follows:
\begin{enumerate}
\item \emph{\label{enu:dps}Degenerate and partly screened}. For $\alpha_{\chi}\lesssim m_{\phi}/m_{\chi}$,
DM will collapse from a degenerate state and 
the partly screened condition $r_{j}<1/m_{\phi}$ is satisfied during the collapse.
The critical number to induce such a collapse is given by
\begin{equation}
N_{{\rm crit,dps}}\approx1.1\times10^{25}\,\alpha_{\chi}^{-6}\left(\frac{m_{\phi}}{{\rm MeV}}\right)^{12}\left(\frac{m_{\chi}}{{\rm GeV}}\right)^{-9}.\label{eq:Ncrit_dps}
\end{equation}
\item \emph{\label{enu:dss}Degenerate and strongly screened.} For $\alpha_{\chi}\gtrsim m_{\phi}/m_{\chi}$,
DM will also collapse from a degenerate state 
but the strongly screened condition, 
$r_{j}>1/m_{\phi}$,  is satisfied. In this scenario, the
analytical expression for the critical DM number cannot be obtained.
One must solve Eq.~(\ref{eq:virial_eq}) numerically under strongly
screened limit. Hence, the virial equation is approximated as
\begin{equation}
\frac{(3\pi^{2})^{2/3}m_{\phi}^{2}}{m_{\chi}y^{2}}=\frac{(4\pi/3)^{1/3}GN_{\chi}\rho_{b}m_{\chi}y^{2}}{m_{\phi}^{2}}+8\alpha_{\chi}m_{\phi}e^{-y}\left(\frac{1}{y}+1\right),\label{eq:viria_ndss}
\end{equation}
where $y=(4\pi/3)^{1/3}r_{{\rm th,deg}}/N_{\chi}^{1/3}$. The $N_{\chi}$
that satisfies the above identity with $y\gtrsim1$ is the critical
number $N_{{\rm crit,dss}}$ for DM to undergo self-collapse in this phase.
\item \emph{Non-degenerate and strongly screened.} If  $N_{{\rm crit}}$
obtained in cases \ref{enu:dps} and \ref{enu:dss} are smaller than
$10^{36}$, then the collapse starts from the non-degenerate state
instead of the degenerate one. Again, one can only obtain the critical DM number
through solving Eq.~(\ref{eq:virial_eq}) numerically in the strongly
screened limit,
\begin{alignat*}{1}
3T_{\chi} & =\frac{(4\pi/3)^{1/3}GN_{\chi}\rho_{b}m_{\chi}y^{\prime2}}{m_{\phi}^{2}}+\frac{(4\pi/3)^{1/3}GN_{\chi}m_{\chi}^{2}m_{\phi}}{y^{\prime}}+8\alpha_{\chi}m_{\phi}e^{-y^{\prime}}\left(\frac{1}{y^{\prime}}+1\right),
\end{alignat*}
where
\begin{equation}
y^{\prime}=\left(\frac{4\pi}{3}\right)^{1/3} m_\phi\frac{r_{{\rm th,nondeg}}}{N_{\chi}^{1/3}}\approx21\,\left(\frac{m_{\phi}}{{\rm MeV}}\right)\left(\frac{m_{\chi}}{{\rm GeV}}\right)^{-1/2}\label{eq:y_prime}.
\end{equation}
Once $N_{{\rm crit,nss}}$ is numerically obtained, there are two
separate scenarios as follows:
\begin{enumerate}
\item If the corresponding $N_{{\rm crit,nss}}$ leads to $y^{\prime}<1$,
the collapse continues without being halted by the Fermi pressure
after DM becoming degenerate when
\begin{equation}
\alpha_{\chi}>0.016\,\left(\frac{m_{\phi}}{{\rm MeV}}\right)^{2}\left(\frac{m_{\chi}}{{\rm GeV}}\right)^{-3/2}.
\end{equation}
\item On the other hand, if $y^{\prime}>1$ for the given $N_{{\rm crit,nss}}$,
then the collapse continues for
\begin{equation}
\alpha_{\chi}>2.7\times10^{-6}\frac{e^{21\left(\frac{m_{\phi}}{{\rm MeV}}\right)\sqrt{\frac{{\rm GeV}}{m_{\chi}}}}}{\left(\frac{m_{\phi}}{{\rm MeV}}\right)\left[1+0.047\left(\frac{{\rm MeV}}{m_{\phi}}\right)\sqrt{\frac{m_{\chi}}{{\rm GeV}}}\right]}.\label{eq:a_cond_ndss}
\end{equation}
\end{enumerate}
\item \emph{Non-degenerate and partly screened.}
When DM is partly screened, we have
	 $y^\prime=1.6m_\phi r/N_\chi^{1/3} \lesssim 1$ and for the non-degenerate case $r$ is
	 determined by $r_{\rm th,nondeg}$. This gives the first criterion
\begin{equation}
N_{\chi,{\rm ndps}}\gtrsim 10^{40} \left(\frac{m_\phi}{\rm MeV}\right)^3 \left(\frac{m_\chi}{\rm GeV}\right)^{-3/2}\label{eq:Nx_ndps}
\end{equation}
However, to trigger gravitational instability, by solving Eq.~(\ref{eq:virial_eq}) with the
Yukawa potential term replaced by $4\pi \alpha_\chi m_\phi/y^{\prime 3}$ in the partly screened
limit \cite{Bramante:2013nma}, the critical number is estimated as
\begin{equation}
N_{{\rm crit,ndps}}\approx2\times10^{34}\,\alpha_{\chi}^{-1}\left(\frac{m_{\phi}}{{\rm MeV}}\right)^{2}\left(\frac{m_{\chi}}{{\rm GeV}}\right)^{-3/2}.\label{eq:Ncrit_ndps}
\end{equation}
Therefore, for DM to collapse into non-degenerate and partly screened phase,
$N_{\chi,{\rm ndps}} > N_{{\rm crit,ndps}}$ or
\begin{equation}
\alpha_\chi \left(\frac{m_\phi}{\rm MeV}\right)\lesssim 2\times 10^{-6} \label{eq:ndps_crit}
\end{equation}
as well as $N_\chi > N_{{\rm crit,ndps}}$ should both be satisfied simultaneously.
However, in our interested  parameter space, both conditions are usually not met simultaneously.   
Even both are satisfied, the condition for collapse from non-degenerate and strongly screened phase
is achieved first. 
So we will not discuss further this scenario.  
A similar statement is also made in Ref.~\cite{Bramante:2013nma} and we refer the interested readers to
this article for details.
\end{enumerate}

From the above discussions, we summarize that the proportional relation that describes $N_{\rm crit}$ in the partly screened limit, Eqs.~(\ref{eq:Ncrit_dps}) and ~(\ref{eq:Ncrit_ndps}), can be approximated as
\begin{equation}
N_{\rm crit} \propto \alpha_\chi^{-a} m_\phi^{b} m_\chi^{-c}, \label{eq:N_crit_behave}
\end{equation}
where $a,b$ and $c$ are positive real numbers.
For strongly screened limit, the Yukawa term has a non-linear dependency on $e^{-y}\sim e^{-1/N_\chi^{1/3}}$. An explicit relation like Eq.~(\ref{eq:N_crit_behave}) 
cannot be obtained. However, we found that Eq.~(\ref{eq:N_crit_behave}) is still a good approximation in this limit although 
it is less sensitive to $\alpha_\chi$, or $a\to 0$. A quantitative justification of this is given in Appendix \ref{sec:prop_relation}.

\subsection{Criteria for continuing collapse}

For each of the four cases discussed above, DMs after the collapse are always
degenerate and relativistic prior to forming a BH. To overcome
the relativistic Fermi pressure that tends to halt the collapse, the condition
\begin{equation}
\alpha_{\chi}>4.7\left(\frac{m_{\phi}}{m_{\chi}}\right)^{2}
\end{equation}
must hold. In addition, the BH 
must
be heavier than $3.4\times10^{36}\,{\rm GeV}$ to prevent itself from evaporation due to Hawking radiation \cite{Bramante:2013nma,Garani:2018kkd}.
Hence the condition 
\begin{equation}
N_{{\rm crit}}>3.4\times10^{36}\,\left(\frac{m_{\chi}}{{\rm GeV}}\right)^{-1}\label{eq:BH_evap}
\end{equation}
should be satisfied in order to destroy the host star within $10^{-3}\,{\rm Gyr}$. 
The content in this section describes a systematic way to determine
the collapse of DM in NS. A more compact and clear steps are summarized
in Table I of Ref.~\cite{Bramante:2013nma}.

\section{Neutron star sensitivities\label{sec:DM_isospin}}

Here we briefly describe how we obtain the NS sensitivity on DM. 
First, we use the method presented in Sec.~\ref{sec:BH_formation} to solve for the critical DM number $N_{\rm crit}$ for given $\alpha_\chi$, $m_\chi$, and $m_\phi$.
As we have observed nearby NS older than $t_{\rm age}=5$ Gyrs, we calculate the total
number of DM captured in the star, $N_\chi$, within $t_{\rm age}$. For $N_\chi>N_{\rm crit}$, we further
check if the collapsed DM particles can successfully form a BH without evaporation and consume the entire star. 
If this happens for a given set of $\alpha_\chi$, $m_\chi$, and $m_\phi$, then these parameter values are excluded with the observation of NS older than $5$ Gyrs.
However, the parameter set remains allowed if it does not lead to BH formation.
\subsection{Exclusion plot over $m_\chi-m_\phi$ plane}
\begin{figure}
	\includegraphics[width=0.45\textwidth]{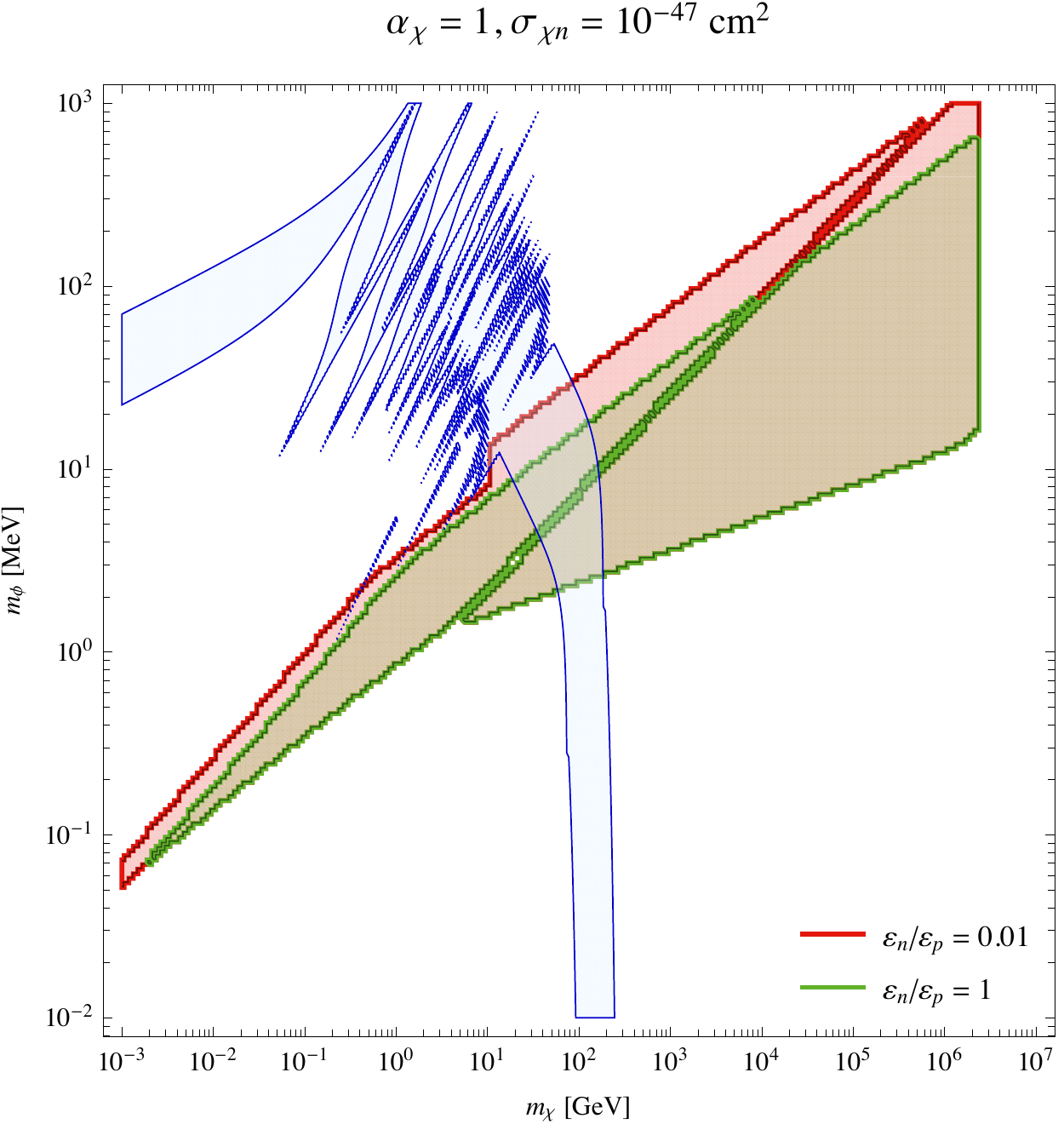}\quad
	\includegraphics[width=0.45\textwidth]{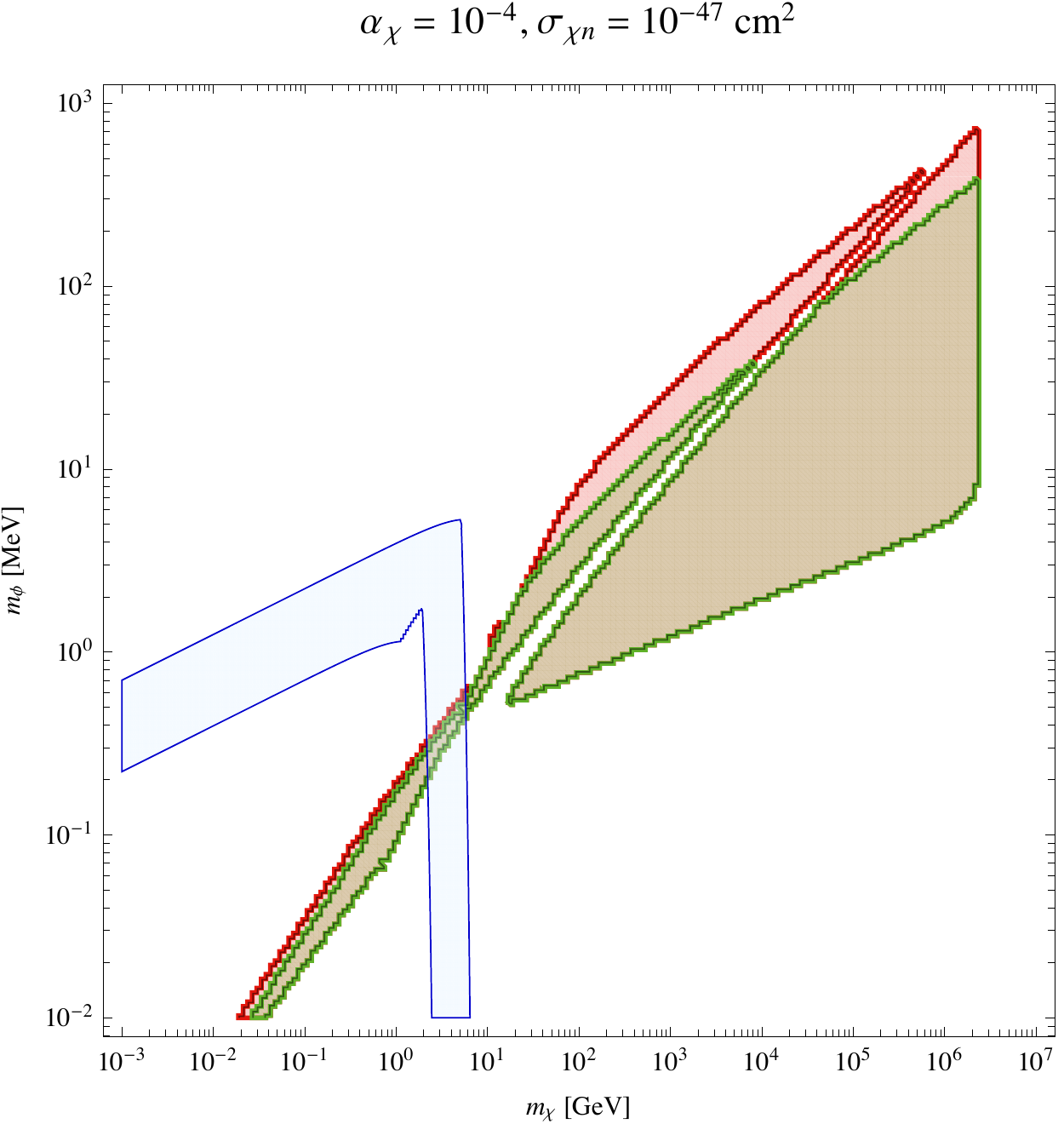}
	
	\caption{\label{fig:excl_plane}The exclusion plane over $m_{\chi}-m_{\phi}$
		plane with $\alpha_{\chi}=1$ (left) and $10^{-4}$ (right) and $\sigma_{\chi n}=10^{-47}\,{\rm cm}^{2}$.
		Red shaded region is excluded by a 5-Gyr-old NS with $\varepsilon_n/\varepsilon_p=0.01$ and
		green with $\varepsilon_n/\varepsilon_p=1$.
		The DM environment has the local density
		$\rho_{\chi}=0.3\,{\rm GeV}\,{\rm cm}^{-3}$ and the velocity dispersion
		$\bar{v}=220\,{\rm km}\,{\rm s}^{-1}$.
		Light blue shaded region is allowed by the SIDM constraint, Eq.~(\ref{eq:sidm_constr}).}
\end{figure}
For convenience, we shall always present $\sigma_{\chi n}$ sensitivity explicitly and the sensitivity to $\sigma_{\chi p}$ can be
inferred from the scaling factor $(\varepsilon_n/\varepsilon_p)^{-2}$.
We present in Fig.~\ref{fig:excl_plane} the exclusion plots over $m_{\chi}-m_{\phi}$
plane with $\alpha_{\chi}=1$ and $10^{-4}$, respectively.
In these plots, it is assumed that NS can survive for 5 Gyrs without being
consumed by DM-forming BH inside it.
The color shaded regions are excluded for $\varepsilon_n/\varepsilon_p=1$ (green) and $\varepsilon_n/\varepsilon_p=0.01$ (red).
In the case of $\varepsilon_n/\varepsilon_p=0.01$, $\sigma_{\chi p}$ is 10,000 times larger than $\sigma_{\chi n}$ and
the contribution from protons to $C_c$ is significantly larger than that from neutrons.
This increases the captured DM number $N_\chi$ in a given period and explains that the excluded region
for isospin violating case is bigger than that for the isospin symmetric one.

It is easily seen that there is a clear cut when $m_{\chi}\gtrsim10^{6}\,{\rm GeV}$.
This is due to the BH mass $m_{{\rm BH}}\equiv N_{\rm crit} m_\chi$ produced from such heavy DMs is too small to satisfy the inequality Eq.~(\ref{eq:BH_evap})
and the BH will evaporate right after its birth.
As $N_{\rm crit} \propto m_\chi^{-c}$ in Eq.~(\ref{eq:N_crit_behave}), the
heavier $m_\chi$ is, the less $N_{\rm crit}$ becomes. Thus, $N_{\rm crit}$ would eventually be too small to satisfy Eq.~(\ref{eq:BH_evap}).
Discussion on this
will be given in the next subsection. In addition, Fig.~\ref{fig:excl_plane}
reproduces  Fig.~1 of Ref.~\cite{Kouvaris:2011gb} with similar
physical interpretations. However there are slight differences on the
excluded regions because the method used by \cite{Bramante:2013nma} for determining the BH formation is slightly different from the method
used by Ref~\cite{Kouvaris:2011gb}.

In Fig.~\ref{fig:excl_plane}, we also show the SIDM allowed
parameter range given by Eq.~(\ref{eq:sidm_constr}) with the light blue shaded region.
Note that the discontinuities are due to the transitions between Eqs.~(\ref{eq:sigxx_born}-\ref{eq:sigxx_hulthen}). To compute $\sigma_{\chi \chi}$, we have input the DM velocity $v=\bar{v}=220~{\rm km~s}^{-1}$ which is the DM velocity dispersion in the Milky Way. Different $v$ leads to different SIDM allowed regions. Detailed discussions on this are given in  Ref.~\cite{Tulin:2013teo}.

\subsection{Sensitivities on $\sigma_{\chi n}$}
\begin{figure}
\begin{centering}
\includegraphics[width=0.45\textwidth]{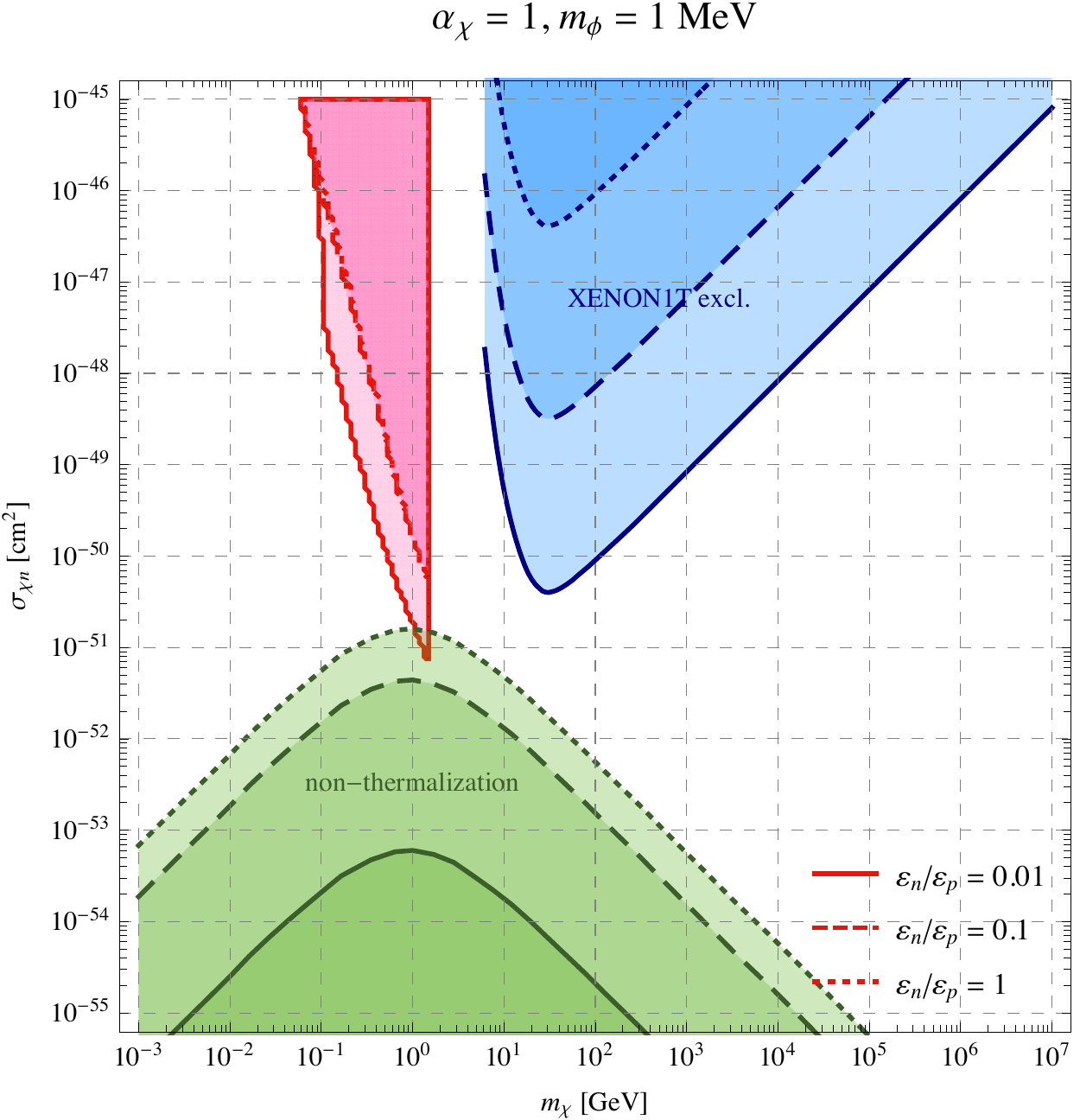}\includegraphics[width=0.45\textwidth]{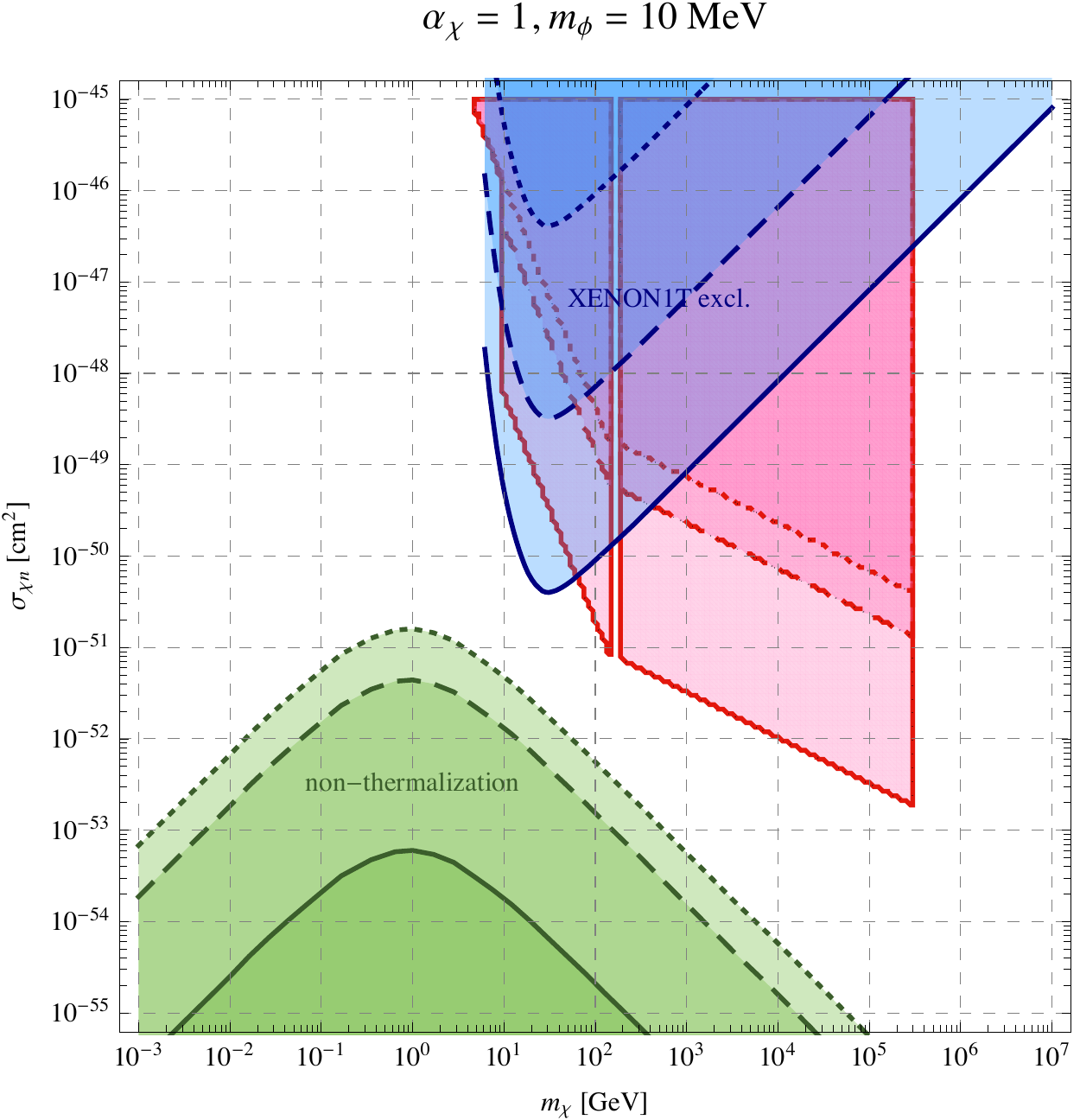}
\par\end{centering}
\begin{centering}
\includegraphics[width=0.45\textwidth]{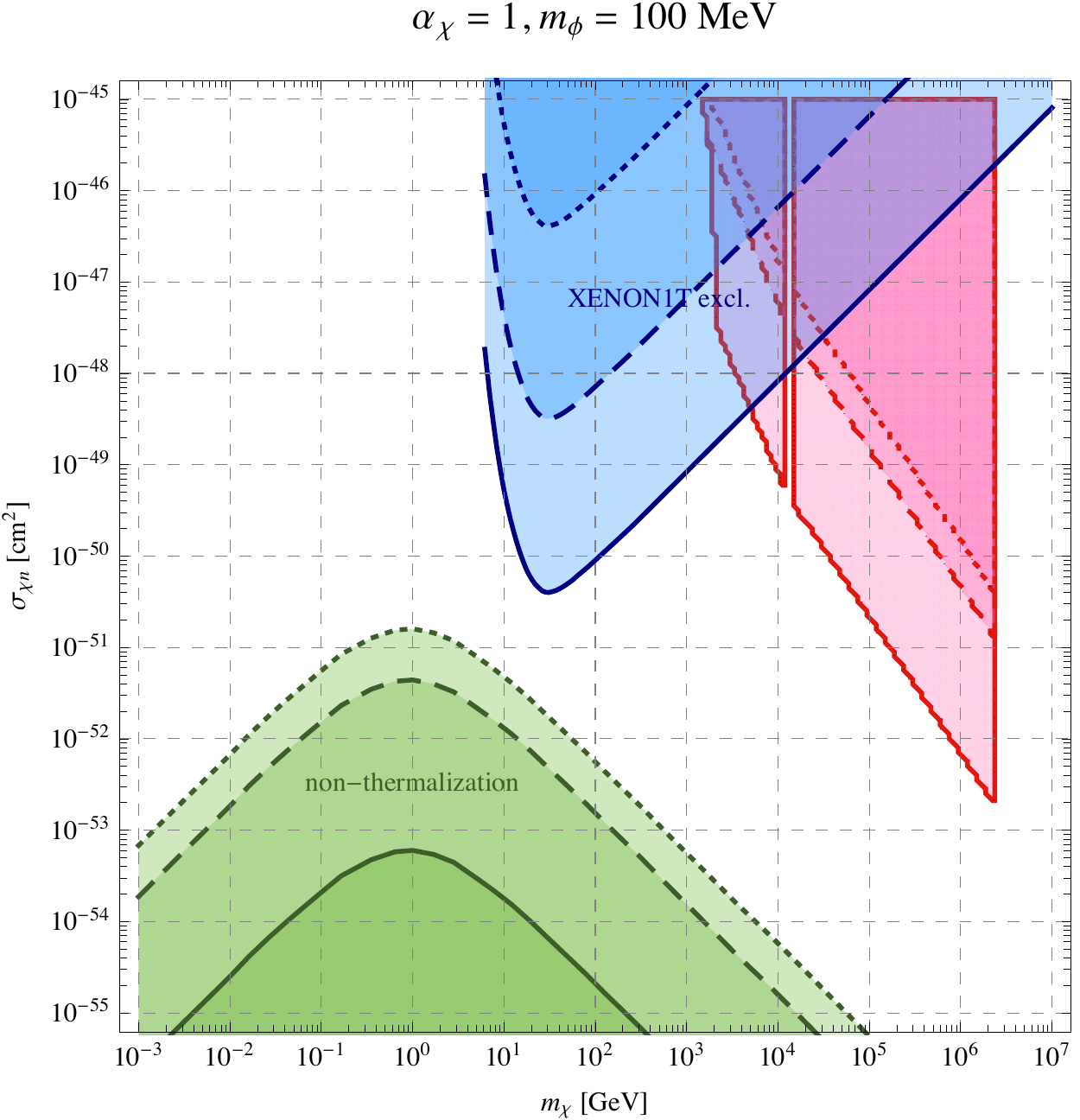}\includegraphics[width=0.45\textwidth]{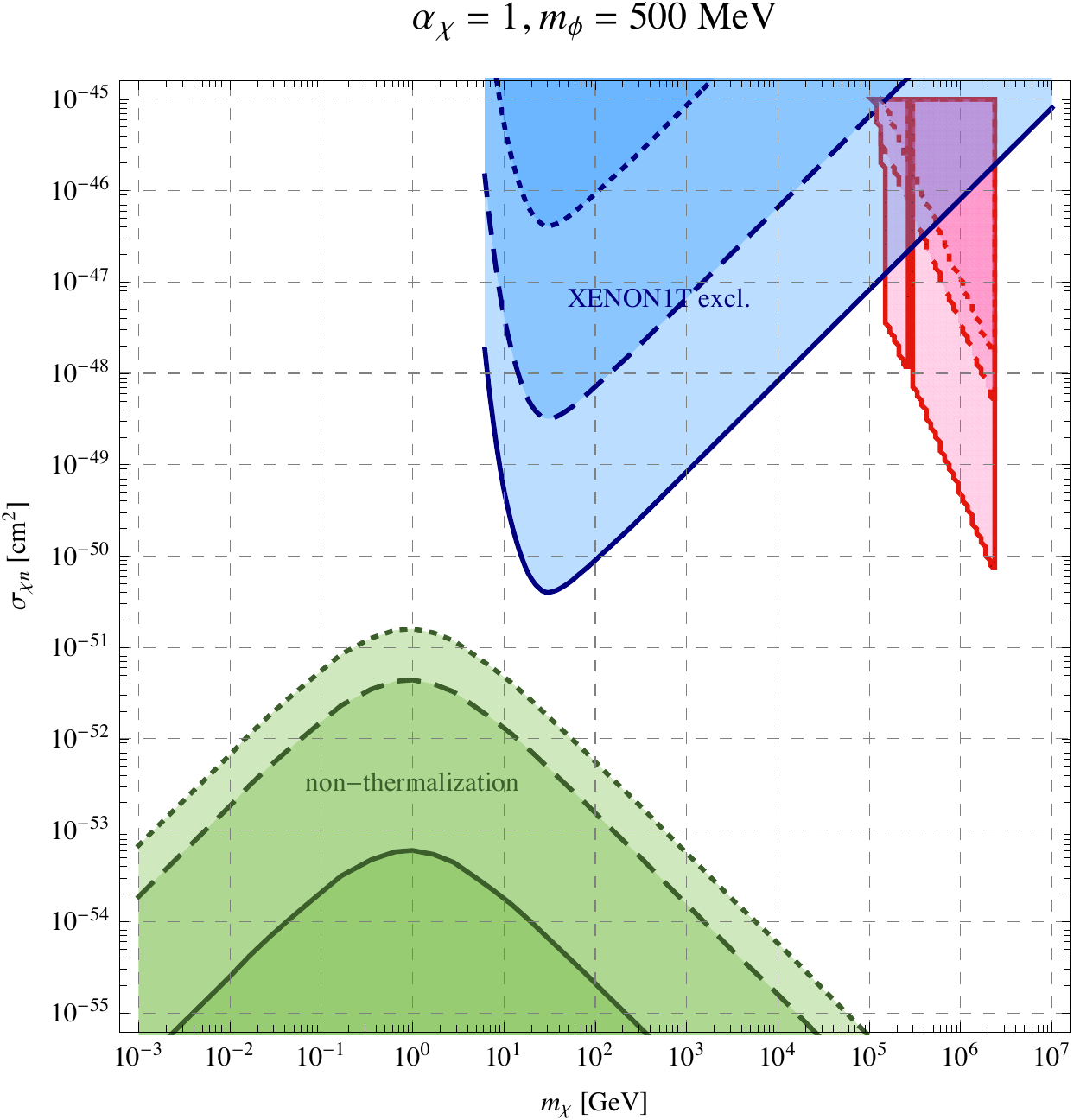}
\par\end{centering}
\caption{\label{fig:a1}The NS sensitivities on $\sigma_{\chi n}$ with $\alpha_{\chi}=1$.
The solid, dashed and dotted lines correspond to $\varepsilon_{n}/\varepsilon_{p}=0.01$,
0.1 and 1, respectively. Pink-shaded regions are excluded by a 5-Gyr-old
NS. Blue-shaded regions are XENON1T exclusions and green-shaded region indicates the parameter range that
DM cannot thermalize with NS within 5 Gyrs.}
\end{figure}

If DM interacts with baryons in the old NS and loses significant kinetic
energy, it would be permanently trapped. 
In principle, we can decompose the capture rate $C_c$ into the contributions
from neutron and proton separately. 
Because DM independently scatter with neutrons and protons, as indicated in Eqs.~(\ref{eq:Cc},\ref{eq:R-}), we have
\begin{equation}
C_{c}\sim C_{c}^{n}\sigma_{\chi n}+C_{c}^{p}\sigma_{\chi p}\label{eq:Cc_dec}
\end{equation}
where $C_{c}^{n,p}$ are the kinematic coefficients due to neutrons
and protons, respectively.\footnote{It is possible to have an approximated form
Eq.~(\ref{eq:Cc_dec}) because we have assumed $\sigma_{\chi b}$ is velocity independent and can be
factored out from the integral in Eq.~(\ref{eq:R-}). In addition, one must bear in mind that $C_c$ has dimension ${\rm s}^{-1}$, thus the
coefficients $C_c^{n,p}$ must carry dimension ${\rm cm}^{-2}\,{\rm s}^{-1}$ to make the unit
correct. However, to our later discussion, one can simply consider $C_c^{n,p}$ as capture related
coefficients regardless their dimensions.}
$N_{\chi}$ resulting from the captures by neutrons and protons are shown separately in Fig.~\ref{fig:Nx}. Generally, the neutron number density is roughly $\mathcal{O}(100)$ times larger than proton's in the star, thus $C_{c}^{n}\gtrsim100C_{c}^{p}$. In
the isospin symmetric case with $\sigma_{\chi n}=\sigma_{\chi p}$, 
the proton capture rate is insignificant and always neglected in 
recent studies. However, in the presence of isospin violation, particularly for
$\varepsilon_{n}/\varepsilon_{p}<1$, i.e., $\sigma_{\chi p}>\sigma_{\chi n}$,  the smaller
proton 
target numbers can be compensated by the larger $\sigma_{\chi p}$.
In this section, we shall discuss the isospin-violating effect on NS sensitivities
to DM-baryon cross section. 
For convenience, we express $\sigma_{\chi p}$ in terms of $\sigma_{\chi n}$ and  $\varepsilon_{n}/\varepsilon_{p}$ as given  by Eq.~(\ref{eq:sigxp}). Hence only NS sensitivities  to $\sigma_{\chi n}$
will be presented with given $\varepsilon_{n}/\varepsilon_{p}$ values.
We take $\varepsilon_{n}/\varepsilon_{p}=\{0.01,0.1,1\}$ as
our benchmark values. For $\varepsilon_{n}/\varepsilon_{p}>1$, we have $\sigma_{\chi n}>\sigma_{\chi p}$ such that
the proton capture rate is always negligible due to $C_{c}^{n}\gg C_{c}^{p}$.
Thus, in the $\varepsilon_{n}/\varepsilon_{p}\to\infty$ limit, the capture rate $C_{c}^{\varepsilon_{n}/\varepsilon_{p}\to\infty}\sim0.98C_{c}^{\varepsilon_{n}/\varepsilon_{p}=1}$.

\begin{figure}
\begin{centering}
\includegraphics[width=0.45\textwidth]{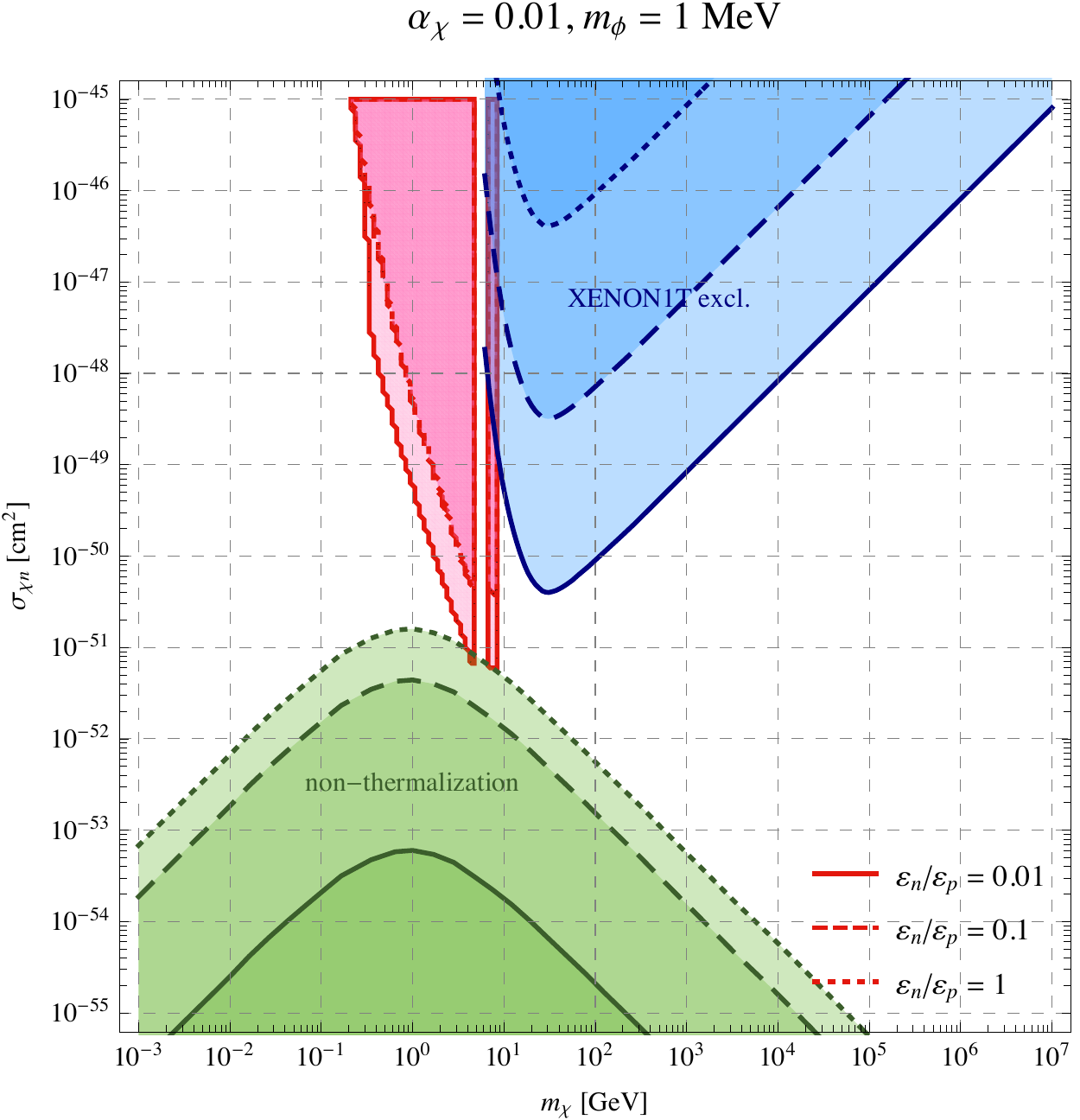}\includegraphics[width=0.45\textwidth]{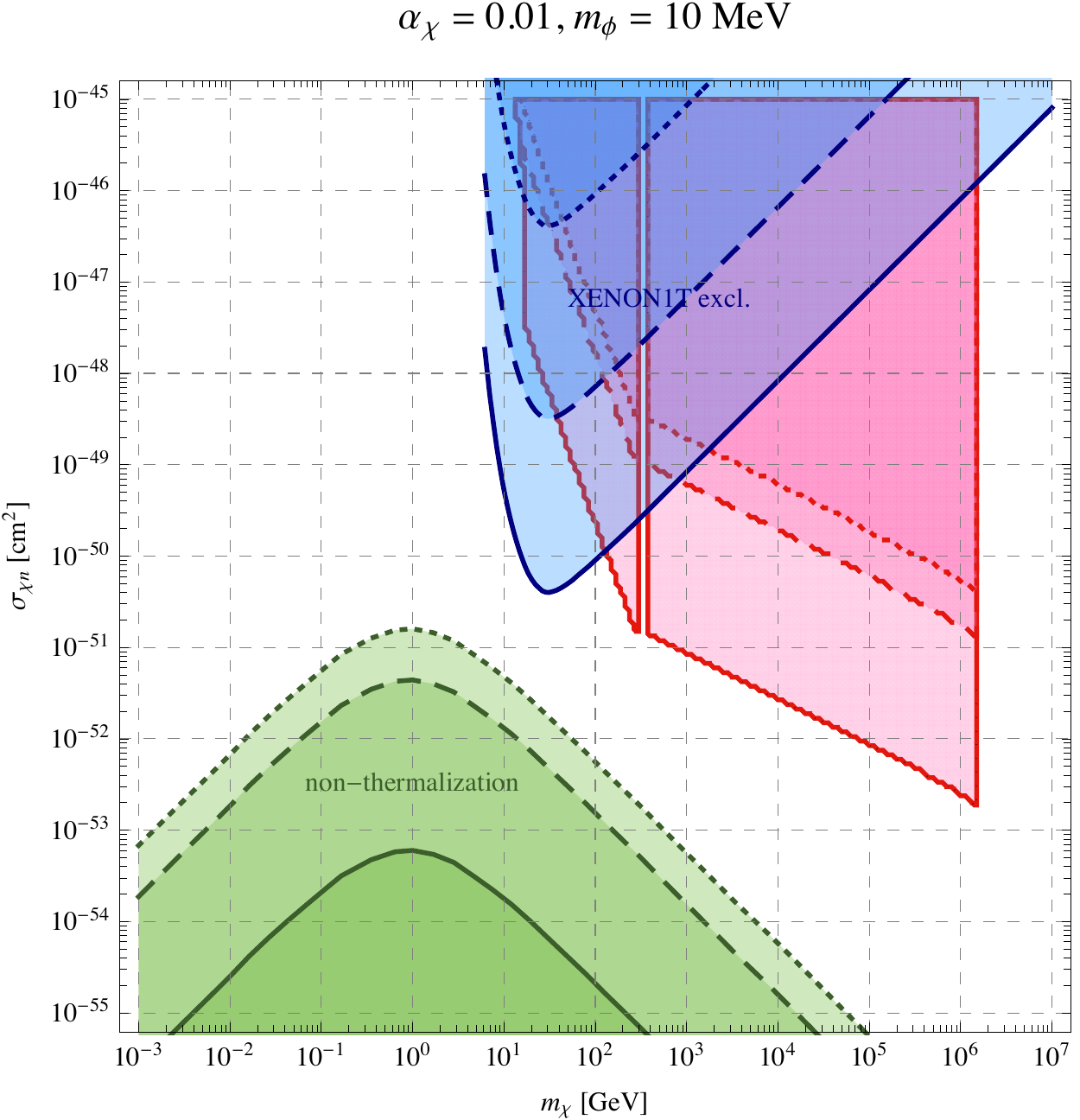}
\par\end{centering}
\begin{centering}
\includegraphics[width=0.45\textwidth]{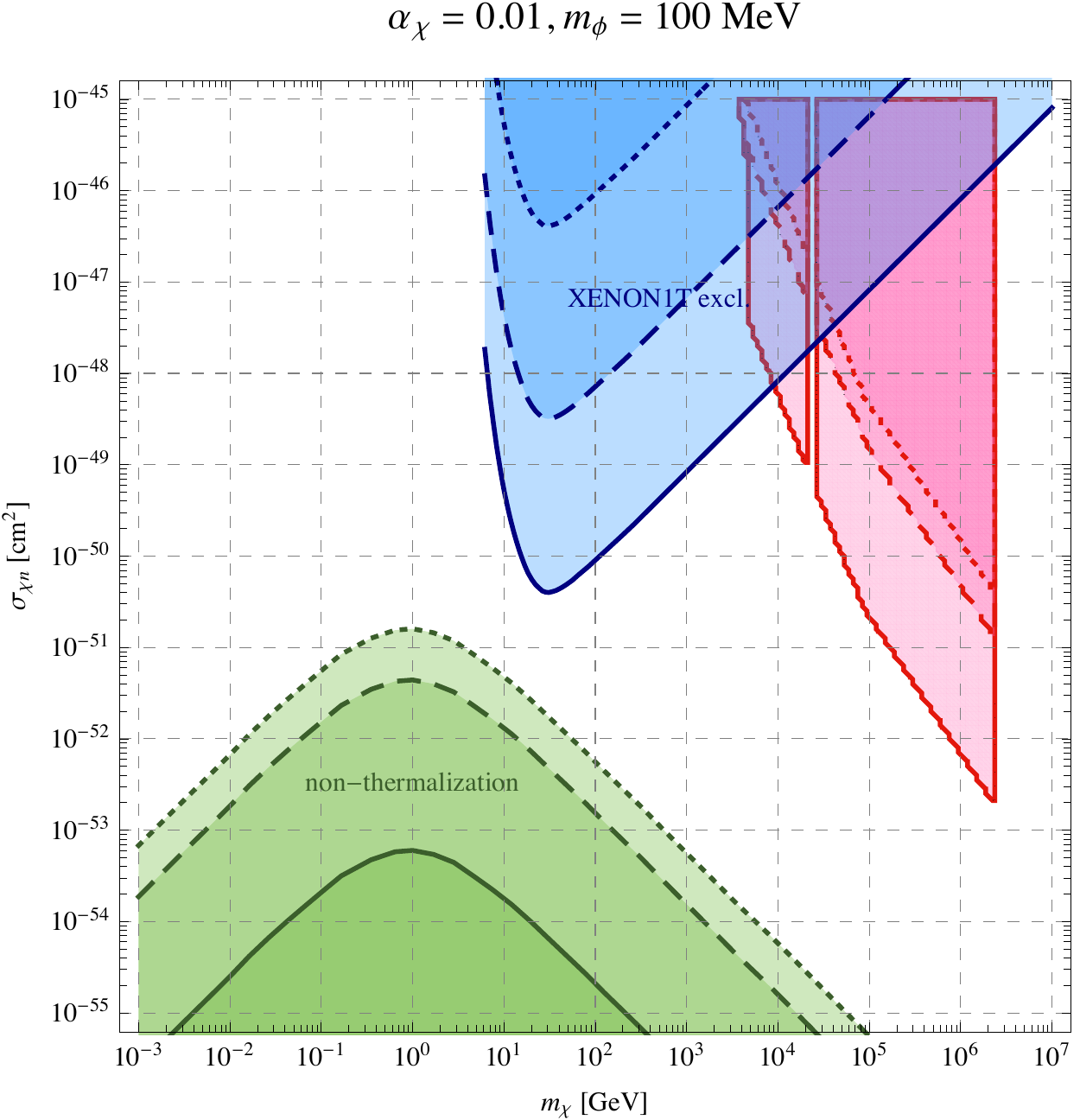}\includegraphics[width=0.45\textwidth]{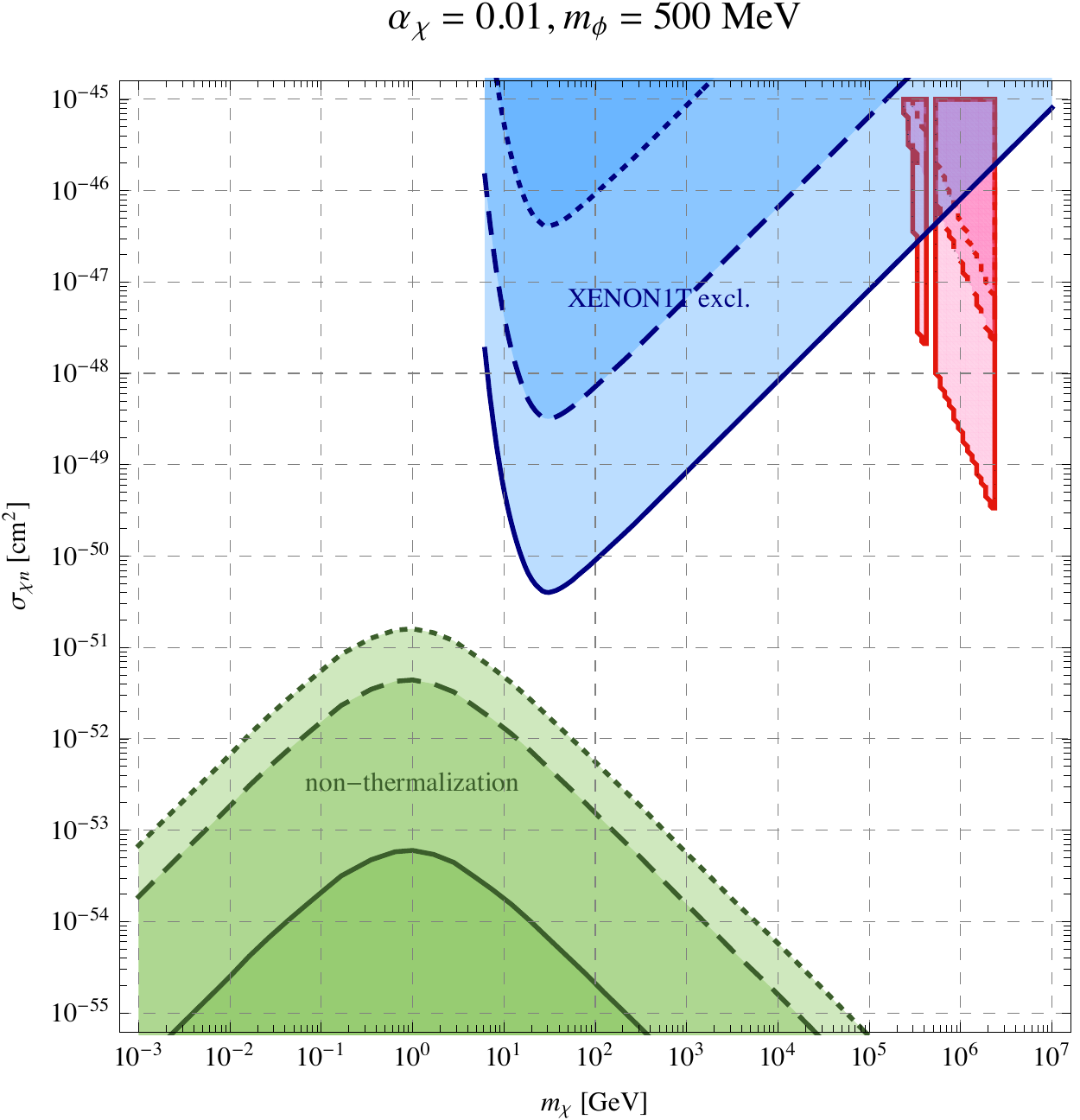}
\par\end{centering}
\caption{\label{fig:a1e-2}Similar to Fig.~\ref{fig:a1}, the NS sensitivities
on $\sigma_{\chi n}$ with $\alpha_{\chi}=0.01$.}
\end{figure}

The 5-Gyr-old NS sensitivities to $\sigma_{\chi n}$ with $\alpha_{\chi}=1$ are shown in Fig.~\ref{fig:a1}.
The environment
has the local DM density $\rho_{\chi}=0.3\,{\rm GeV}\,{\rm cm}^{-3}$
and the DM velocity dispersion $\bar{v}\approx220\,{\rm km}\,{\rm s}^{-1}$
which are similar to the Solar System. The solid, dashed and dotted
lines correspond to $\varepsilon_{n}/\varepsilon_{p}=0.01$, 0.1 and 1, respectively. 
	Pink-shaded region is the parameter range that DM can form BH inside the star and consume the star entirely. The observation of NS older than 5 Gyrs thus rules out this region.
Blue-shaded regions are XENON1T
exclusions scaled by $\varepsilon_{n}/\varepsilon_{p}$ according to
Eq.~(\ref{eq:Fz}). 
	Green-shaded region indicates the parameter range that $\sigma_{\chi n}$ is too small to thermalize with NS for a given $m_{\chi}$. Thus $T_{\chi}\neq T_{{\rm NS}}$. 
If DM has higher temperature, it gains additional thermal pressure
to counteract the gravitational potential and the collapse is more difficult
to happen. Therefore,  NS sensitivities are not reliable in this
region since all studies so far assume $T_{\chi}=T_{{\rm NS}}=10^{5}\,{\rm K}$.
The non-thermalized regions for different $\varepsilon_{n}/\varepsilon_{p}$ are quantitatively
scaled from $\varepsilon_{n}/\varepsilon_{p}=1$ using the result in Ref.~\cite{Garani:2018kkd} according
to the interaction length $\lambda_{{\rm int}}^{-1}=n_{b}\sigma_{\chi b}$.
However, since DM thermalization with NS medium
is a complex topic and off the scope of this work, we refer the interested
reader to Ref.~\cite{Bertoni:2013bsa} for details.

\begin{figure}
\begin{centering}
\includegraphics[width=0.45\textwidth]{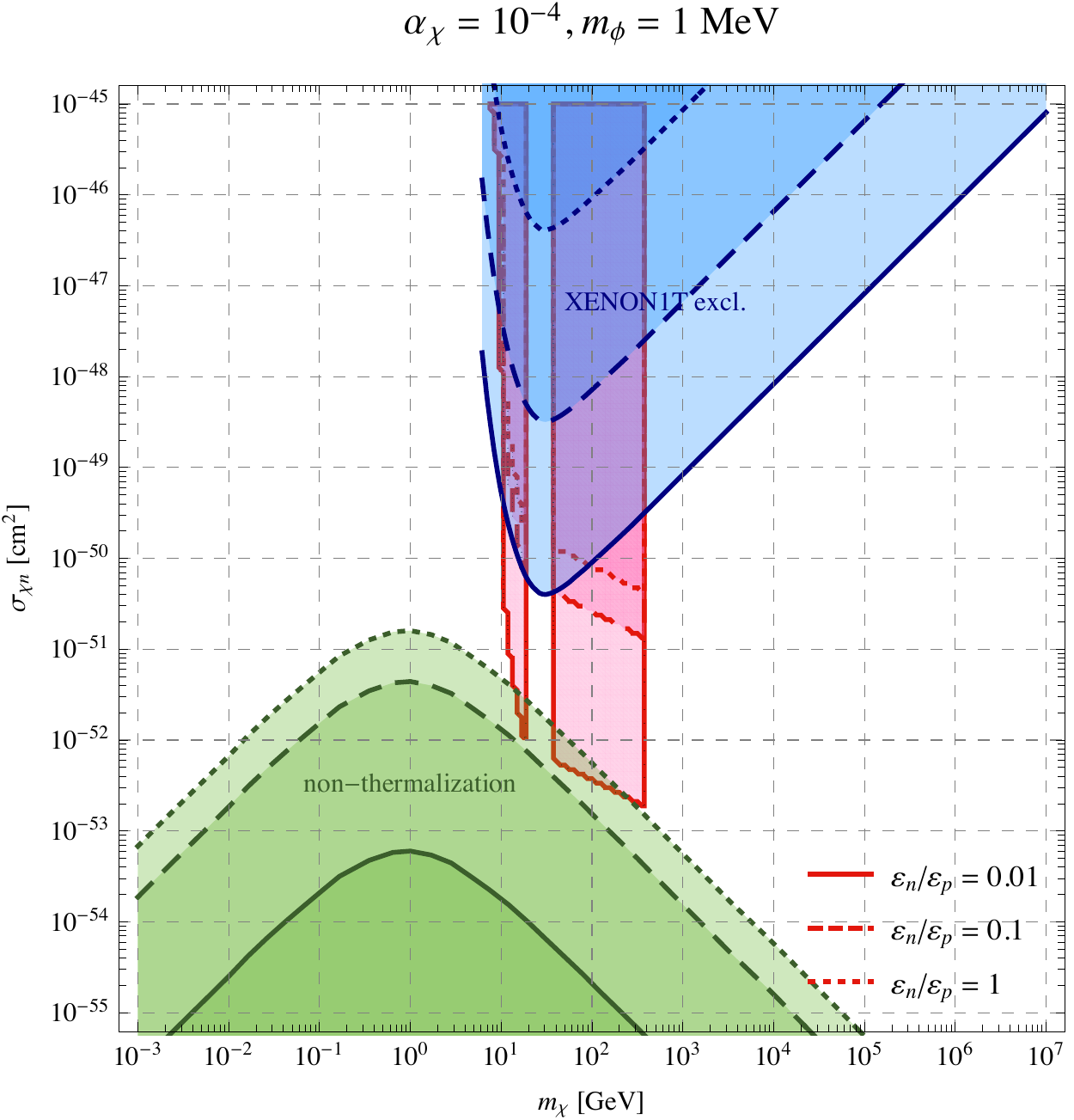}\includegraphics[width=0.45\textwidth]{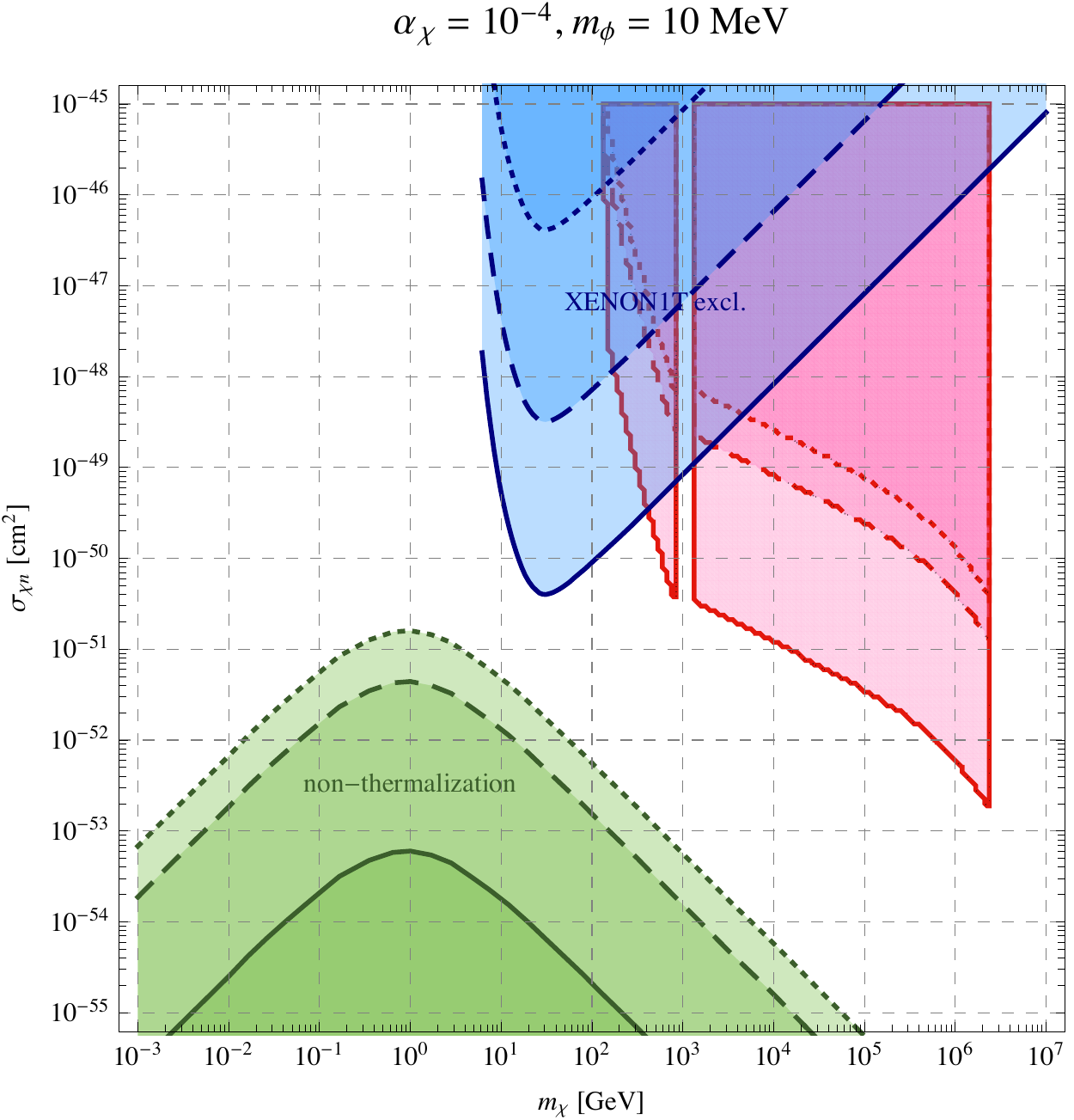}
\par\end{centering}
\begin{centering}
\includegraphics[width=0.45\textwidth]{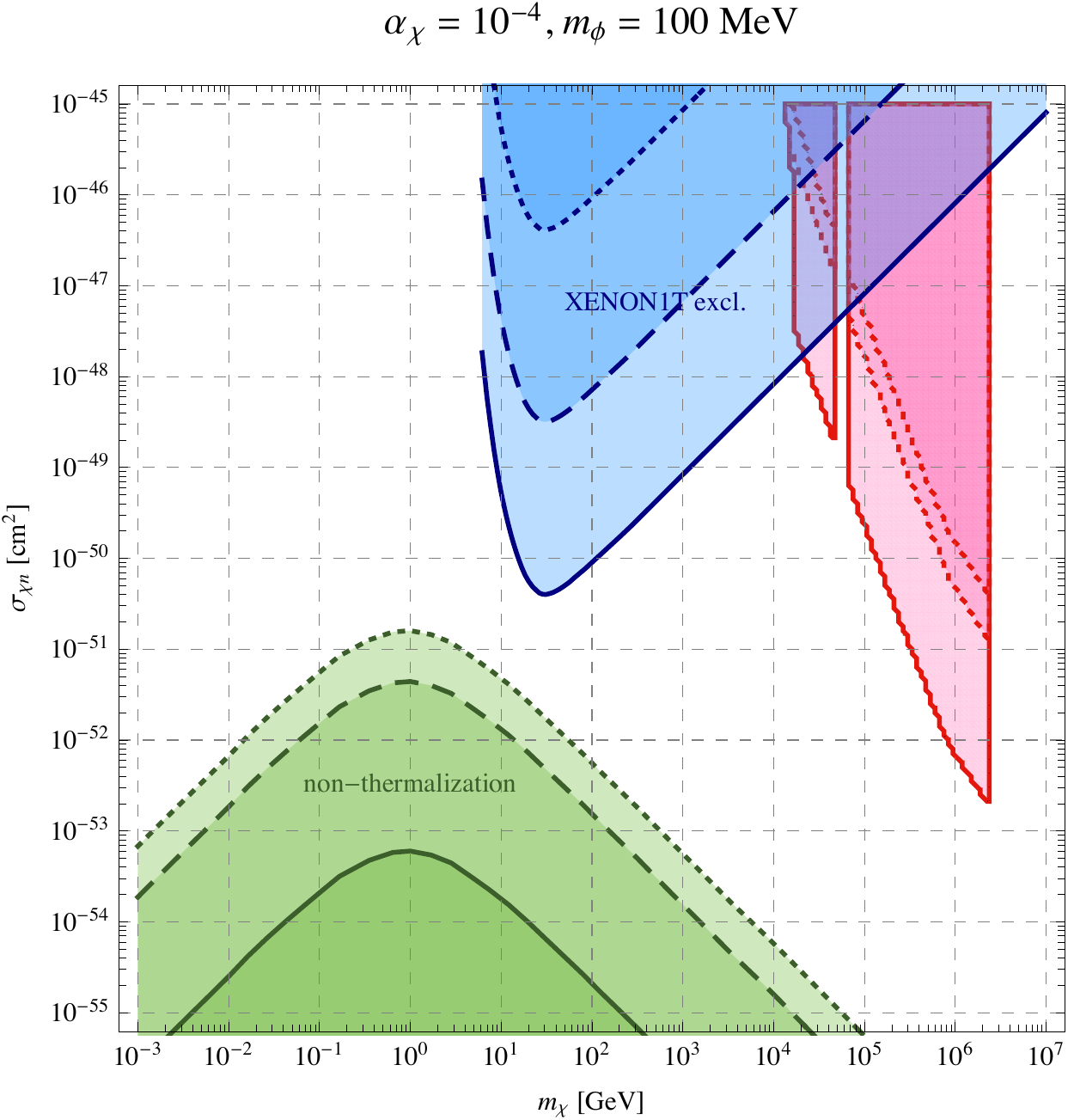}\includegraphics[width=0.45\textwidth]{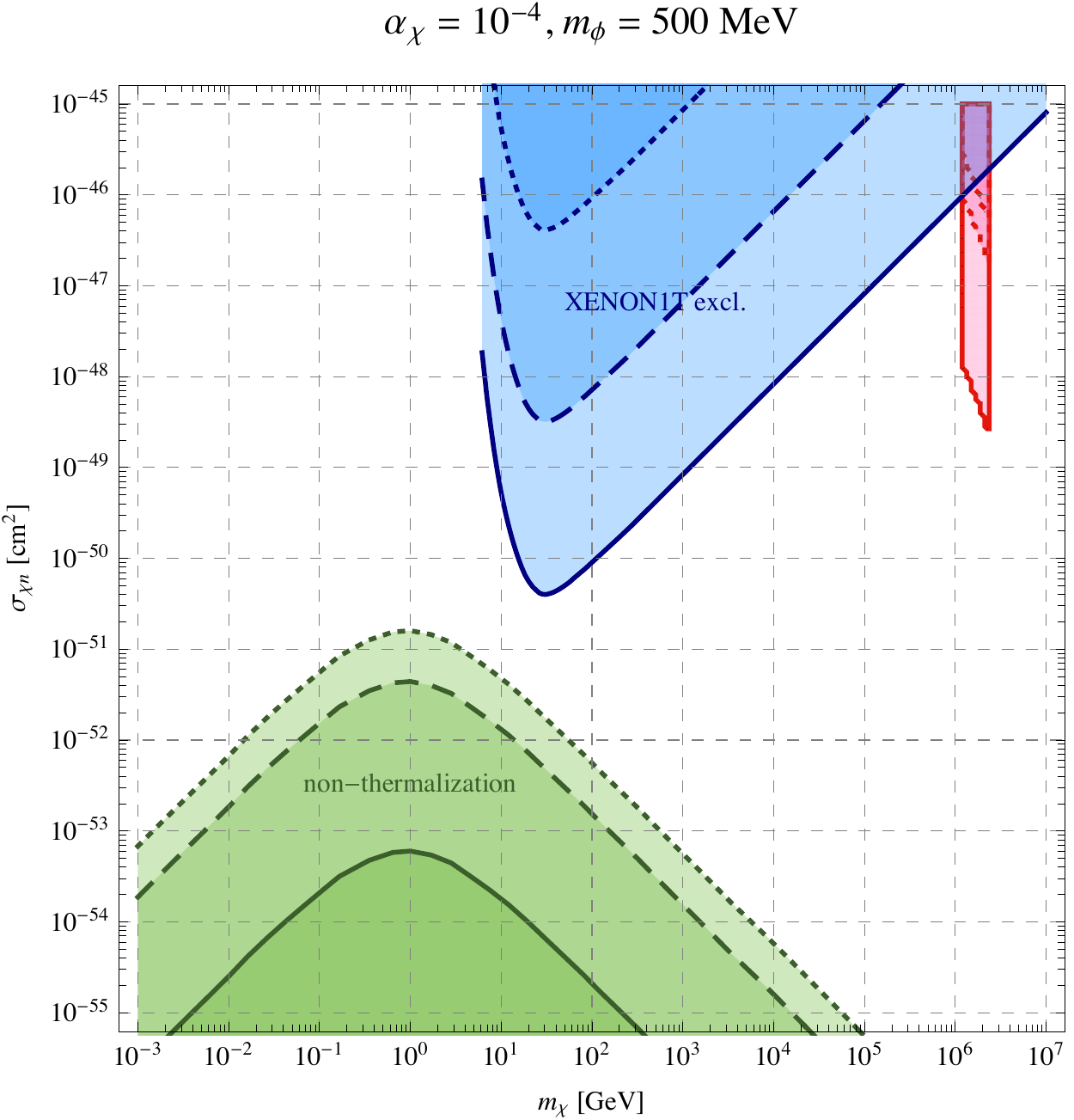}
\par\end{centering}
\caption{\label{fig:a1-4}Similar to Fig.~\ref{fig:a1}, the NS sensitivities
on $\sigma_{\chi n}$ with $\alpha_{\chi}=10^{-4}$.}
\end{figure}
From Fig.~\ref{fig:a1}, 
the excluded parameter space is shifted to the heavier $m_{\chi}$ region as $m_{\phi}$ becomes
larger.
This behavior can be understood as follows.
From Eq.~(\ref{eq:N_crit_behave}), we found that $N_{\rm crit}$ is related to model parameters such that
\begin{equation}
N_{{\rm crit}}\propto\alpha_{\chi}^{-a}m_{\phi}^{b}m_{\chi}^{-c},\label{eq:Ncrit_approx}
\end{equation}
where $a,b,c$ are some positive real numbers larger than one, but
the true values depend on the four phases in Sec.~\ref{subsec:after_coll}.
To initiate the gravitational collapse, 
	it is imperative to have sufficient DM particles captured inside the star.
Since the
capture rate $C_{c}\propto m_{\chi}^{-1}$, we have the condition for
the collapse:
\begin{equation}
\frac{N_{{\rm crit}}}{N_{\chi}}\sim\frac{N_{{\rm crit}}}{C_{c}t_{{\rm age}}}\sim\alpha_{\chi}^{-a}m_{\phi}^{b}m_{\chi}^{-c+1}=\alpha_{\chi}^{-a}m_{\phi}^{b}m_{\chi}^{-c^{\prime}}=1.\label{eq:NN}
\end{equation}
In the above, we neglect the constant NS age $t_{{\rm age}}$ in
the final step.  We also note that $c^{\prime}>0$ in general.
	For a larger $m_{\phi}$, the collapse also requires a larger $m_{\chi}$ to ensure the validity of Eq.~(\ref{eq:NN}).
This explains why a larger $m_{\phi}$  leads to an excluded parameter space in the heavier DM region.

It is easily seen that the sensitivities to $\sigma_{\chi n}$ become more stringent
when $\varepsilon_{n}/\varepsilon_{p}$ is smaller. 
This is due to the enhancement of $\sigma_{\chi p}$ versus $\sigma_{\chi n}$.
Thus,
taking $\varepsilon_{n}/\varepsilon_{p}=0.1$ for instance, the corresponding $\sigma_{\chi p}$
is $100$ times larger than $\sigma_{\chi n}$. 
	This makes the fraction of proton contributions to the capture rate comparable to that of neutron contributions.
The capture rate in this scenario is roughly
estimated as $C_{c}^{\varepsilon_{n}/\varepsilon_{p}=0.1}\approx2C_{c}^{\varepsilon_{n}/\varepsilon_{p}=1}$
based on Eq.~(\ref{eq:Cc_dec}). For $\varepsilon_{n}/\varepsilon_{p}=0.01$, we have
$C_{c}^{\varepsilon_{n}/\varepsilon_{p}=0.01}\approx100C_{c}^{\varepsilon_{n}/\varepsilon_{p}=1}$. This explains
why a smaller $\varepsilon_{n}/\varepsilon_{p}$ 
leads to a better sensitivity to $\sigma_{\chi n}$.

Other sensitivities for $\alpha_{\chi}=0.01$ and $10^{-4}$ are also
shown in Figs.~\ref{fig:a1e-2} and \ref{fig:a1-4}, respectively.
Note that for a fixed $m_{\phi}$, smaller $\alpha_{\chi}$ leads to an excluded region in the heavier DM range.
This can also be understood
from Eq.~(\ref{eq:NN}). For smaller $\alpha_{\chi}$, $\alpha_{\chi}^{-a}$
is in fact enhanced. To maintain the validity of  Eq.~(\ref{eq:NN}), $m_{\chi}$ should
larger.

We note that there is a sharp boundary on the right of the excluded region in each plot.
	It was already mentioned in the last subsection that the BH mass $m_{{\rm BH}}\equiv N_{{\rm crit}}m_{\chi}$ formed after collapse could be too small to overcome the evaporation effect. 
	Since $N_{\rm crit}\propto m_{\chi}^{-c}$ with $c>1$ according to Eq.~(\ref{eq:NN}), we have  $m_{{\rm BH}}\propto m_{\chi}^{-c+1}$.  
Hence, beyond a critical $m_{\chi}$,  the produced BH will evaporate
right after its birth, i.e., the star will not be consumed by such DM-forming
BH.
Besides the sharp boundaries, there are gaps between the low and high $m_{\chi}$ regimes of
the excluded regions. It is explained in Ref.~\cite{Bramante:2013nma} that the algorithm used to
solve the virial equation treats the collapse of DM particles either all
from degenerate or non-degenerate state depending on 
the corresponding $N_{{\rm crit}}$.
$N_{{\rm crit}}$ for low $m_{\chi}$ is greater than that for high $m_{\chi}$.
Thus, the collapse usually happens
in a degenerate phase for low $m_{\chi}$ and non-degenerate phase for high
$m_{\chi}$. 
	The gap occurs at the transition from degenerate
to non-degenerate phases after collapse, but the condition (\ref{eq:a_cond_ndss}) is not satisfied.
This is believed to be an 
artifact of the algorithm
due to the insufficient consideration of the subtle effect in the
transition region. This artifact can be fixed with the gap closed provided the full dynamics of collapse could be simulated as proposed
in Ref.~\cite{Bramante:2013nma}. 
We leave this for future studies since it is out of the scope of the current work. 
\section{Summary\label{sec:Summary}}

In this work, we 
have pointed out that isospin violation could
significantly change the NS sensitivities to $\sigma_{\chi n}$. 
Up to recent NS studies, proton contribution to the capture rate is ignored due to the assumption of isospin symmetry 
in the fermionic DM-baryon interaction.\footnote{For bosonic DM with isospin violation
in the NS is studied in Ref.~\cite{Zheng:2014fya}.}
On the other hand, we have demonstrated that the
proton contribution to the capture rate becomes comparable and even
exceeds the neutron contribution for $\varepsilon_{n}/\varepsilon_{p}$ much smaller than unity. Thus the sensitivity regions are enlarged.
This has not been taken into
account for the case of attractive fermionic DM. In our new study,
the DM capture rate is based on a more realistic NS profile. Moreover, the
suppression due to Pauli blocking effect has been considered in calculating $C_c$. 

For the DM-forming BH inside the NS, the fermionic case contains very rich
dynamics as we have summarized in Sec.~\ref{sec:BH_formation}.
A systematic algorithm to compute the fate of NS based on the DM particle
parameters $\alpha_{\chi}$, $m_{\chi}$, $m_{\phi}$ and the captured
$N_{\chi}$ in the NS is given in Ref.~\cite{Bramante:2013nma}. Here we not only
implement it into our study, we also provide a python package \texttt{dm2nsbh}
accompanied with this work. This package is a realization of the above algorithm 
and released
on the github. In \texttt{dm2nsbh}, functions for calculating capture
rate $C_{c}$ and DM number $N_{\chi}$ with different $\varepsilon_{n}/\varepsilon_{p}$
are also provided. We hope this code release could stimulate
a more thorough and improved studies on this issue.
In addition, the artifact in the algorithm mentioned in Sec.~\ref{sec:BH_formation}
and Ref.~\cite{Bramante:2013nma} that causes the gaps between low and high $m_{\chi}$ regimes
in the plots can be resolved eventually.

In closing, we stress that the microscopic feature of DM self-interaction
is not yet well understood at the present time. Thus, if DM self-interaction
is repulsive instead of attractive, the last term in the virial equation, Eq.~(\ref{eq:virial_eq}), changes sign and consequently counteracts the gravitational attraction. 
This effect would disfavor the collapse of DMs into a BH. Hence NS is  no longer able to constrain DM parameter space.  
On the other hand, if DMs attract each other as studied in this work, 
it is possible for them to form a 
large nugget that contains enormous number of DM particles as investigated in Refs.~\cite{Wise:2014jva,Gresham:2017zqi,Gresham:2017cvl,Gresham:2018anj}.
The nugget may have different scattering cross section with the baryons
in NS.
Hence the method employed here for constraining DM properties should be modified. We relegate this to future studies. 

\appendix
\section{Model details}\label{sec:model_detail}
In this appendix we present some details of the phenomenological framework given in Sec.~\ref{sec:model}. The purpose is to understand how DM and its self-interaction are introduced in this work and the requirement for an attractive DM self-interaction.

To begin with, we introduce two Dirac fermions $\xi$, $\eta$ and the complex scalar field $\Phi$ in the dark sector. Both $\xi$ and $\Phi$ carry $U(1)_d$ charge $g_d$ while
$\eta$ is a $U(1)_d$ singlet. Thus, we can write down the following mass related Lagrangian: 
\begin{equation}
\mathcal{L}_M =-m_\xi \bar{\xi} \xi - m_\eta \bar{\eta}\eta - y\Phi^*\bar{\xi} \eta - y\Phi \bar{\eta} \xi,
\label{eq:yukawa}
\end{equation}
where $y$ is the Yukawa coupling. 
Spontaneous symmetry breaking gives $\Phi=(\phi +v_d+i\sigma)/\sqrt{2}$.
Substituting this expression into Eq.~(\ref{eq:yukawa}), we obtain fermionic mass terms as 
\begin{equation}
\mathcal{L}_{M} \supset -m_\xi \bar{\xi} \xi - m_\eta \bar{\eta}\eta - yv_d(\bar{\xi} \eta + \bar{\eta}\xi)/\sqrt{2}.
\end{equation}
To diagonalize the above mass terms, we rotate $(\xi,\eta)$ by an angle $\theta$ and obtain
the physical fields $(\psi,\chi)$ such that $\psi=\xi \cos\theta - \eta\sin\theta$ and $\chi=\xi\sin\theta + \eta\cos\theta$, with
\begin{equation}
\tan2\theta=\frac{\sqrt{2}yv_d}{m_{\eta}-m_{\xi}}.
\end{equation}
The diagonalized mass terms are given by 
\begin{equation}
\mathcal{L}_{m} = -m_\psi \bar{\psi} \psi - m_\chi \bar{\chi}\chi,
\end{equation}
where $m_\psi=M_+$ and $m_\chi=M_-$ with
\begin{equation}
M_\pm = \frac{(m_\xi + m_\eta)\pm \sqrt{(m_\xi - m_\eta)^2+2v_d^2y^2}}{2}.
\end{equation}
We note that $m_\psi > m_\chi$ in general. In the $(\psi,\chi)$ basis, the Yukawa interactions can be written as 
\begin{equation}
\mathcal{L}_{\rm int} = y(\sin 2\theta \bar{\psi}\psi - \sin 2\theta \bar{\chi}\chi
- \cos 2\theta \bar{\psi}\chi - \cos 2\theta\bar{\chi}\psi)\phi/\sqrt{2}. \label{eq:L_int_phys}
\end{equation}
If $m_\psi > m_\chi + m_\phi$, $\psi$ can decay into
$\chi$ and $\phi$ such that $\chi$ is the desired DM candidate.  
Besides, the second term of Eq.~(\ref{eq:L_int_phys}) characterizes the DM self-interaction
mediated by $\phi$. However, $U(1)_d$ boson can also be another mediator of
DM self-interaction through the gauge term $g_v \bar{\chi}\gamma_\mu  Z_d^\mu\chi$
where $g_v = g_d \sin^2\theta$. Let $g_s = -y\sin 2\theta/\sqrt{2}$, the Feynman diagrams correspond to the two contributions are shown in Fig.~\ref{fig:feyn}.

\section{Proportional relation of $N_{\rm crit}$}\label{sec:prop_relation}
In this appendix, we justify the proportional relation, Eq.~(\ref{eq:N_crit_behave}), given in Sec.~\ref{subsec:after_coll}.
The purpose of this appendix is to show the dependencies of $N_{\rm crit}$ on
$\alpha_\chi$, $m_\chi$ and $m_\phi$ in an illustrative way instead of deriving
the analytical expressions for $N_{\rm crit}$. 
Various approximations are employed here.
Those who are interested in obtaining  more accurate results should refer to Sec.~\ref{subsec:after_coll} or directly use the package \texttt{dm2nsbh}.

\subsection{Partly screened limit}
To trigger the gravitational instability, the last two terms on the RHS
of Eq.~(\ref{eq:virial_eq}) should become more important than the NS background potential (the first term).\footnote{
	If the NS background potential always dominates, then it will be canceled by the thermal energy $2E_{k}$. The entire system will never collapse.}
Moreover, when the collapse begins in the partly screened limit, the Yukawa potential (the third term) can be approximated by
the Coulomb potential due to the interparticle distance $r_j<1/m_\phi$. It will also overcome the DM self-gravity (the second term).
Thus, the virial equation becomes
\begin{equation*}
\frac{(3\pi^2)^{2/3}m_\phi^2}{m_\chi y^2}=\frac{4\pi\alpha_\chi m_\phi}{y^3}.
\end{equation*}
With $y=(4\pi/3)^{1/3}m_\phi r_{\rm th,deg}/N_\chi^{1/3}$, we obtain a simple analytical approximation for $N_{\rm crit}$ in the degenerate case,
\begin{equation}
N_{\rm crit,dps}\approx \frac{243}{8192}\sqrt{\frac{3\pi^{7}}{(G\rho_b)^{3}}} \alpha_\chi^{-6} m_\phi^{12}m_\chi^{-6},
\end{equation}
which is the same as Eq.~(\ref{eq:Ncrit_dps}).
On the other hand, for the non-degenerate and partly screened case, the virial equation becomes
\begin{equation*}
-2T_\chi = \frac{4\pi \alpha_\chi m_\phi}{y^{\prime 3}}
\end{equation*}
where $y^\prime$ is given in Eq.~(\ref{eq:y_prime}). Solving this equation yields
\begin{equation}
N_{\rm crit,ndps} =\left(\frac{9T_\chi^{5/3}}{4\pi G \rho_b}\right)^{3/2} \alpha_\chi^{-1} m_\phi^2 m_\chi^{-3/2}
\end{equation}
which is the same as Eq.~(\ref{eq:Ncrit_ndps}).

\subsection{Strongly screened limit}

Here we only present the degenerate case for simplicity. In this limit with
DM self-gravity term ignored, Eq.~(\ref{eq:viria_ndss}) can be recast as
\begin{equation*}
\frac{(3\pi^{2})^{2/3}m_{\phi}}{8\alpha_{\chi} m_{\chi}}=
y( y+1)e^{-y}.
\end{equation*}
NS collapse begins when $y\gtrsim 1$. 
Taking logarithm on both sides of the above equation, the RHS may be approximated as $-y$ for $y\gg 1$. Hence we arrive at
\begin{equation*}
y = -\ln\left[\frac{(3\pi^2)^{2/3}m_\phi}{8\alpha_\chi m_\chi }\right] \equiv -\ln f(\alpha_\chi,m_\phi, m_\chi).
\end{equation*}
Thus, even $\alpha_\chi$, $m_\phi$ and $m_\chi$ can vary over many orders of magnitude, $y$ 
only varies slowly.
Its value is greater than unity so long as $\alpha_\chi m_\chi/m_\phi > (3\pi^2)^{2/3}/8\approx 0.179$. This criterion is always satisfied since the collapse from the strongly screened limit requires $\alpha_\chi m_\chi/m_\phi \gtrsim 1$. Using the relation between $y$ and $N_{\chi}$, we obtain
\begin{equation}
N_{\rm crit,dss} = \frac{16}{3}\frac{\pi^3}{(\ln f)^6}\frac{ m_\phi^6 m_\chi^{-3}}{(4\pi G \rho_b/3)^{3/2}},
\end{equation}
which depends on $\alpha_\chi$ only logarithmically.
Based on the analysis given in this appendix, we argue quantitatively that the proportional relation, Eq.~(\ref{eq:N_crit_behave}),
\begin{equation*}
 N_{\rm crit} \propto \alpha_\chi^{-a} m_\phi^{b} m_\chi^{-c}
\end{equation*}
is reasonable and justified.

\begin{acknowledgments}
YHL thanks the authors of Ref.~\cite{Garani:2018kkd} for providing
the data plot of proton capture rate as well as the kind support by
the Academia Sinica, Taiwan. GLL is supported by the Ministry of Science and
Technology, Taiwan under Grant No.~107-2119-M-009-017-MY3.
YHL thanks Gang Guo and Meng-Ru Wu for useful discussions.
\end{acknowledgments}


\begin{thebibliography}{99}
	
\bibitem{Aad:2015zva}
G.~Aad et al. [ATLAS Collaboration],
Eur.\ Phys.\ J.\ C {\bf 75}, 
299 (2015)
[Erratum ibid {\bf 75}, 
408 (2015)]
[arXiv:1502.01518 [hep-ex]].

\bibitem{Abdallah:2015ter}
J.~Abdallah et al.,
Phys.\ Dark Univ.  {\bf 9-10}, 8 (2015)
[arXiv:1506.03116 [hep-ph]].

\bibitem{Aalbers:2016jon} 
J.~Aalbers {\it et al.} [DARWIN Collaboration],
JCAP {\bf 1611}, 017 (2016)
[arXiv:1606.07001 [astro-ph.IM]].

\bibitem{Akerib:2016vxi}
D.~S.~Akerib et al. [LUX Collaboration],
Phys.\ Rev.\ Lett.  {\bf 118}, 
021303 (2017)
[arXiv:1608.07648 [astro-ph.CO]].

\bibitem{Amole:2017dex} 
C.~Amole et al. [PICO Collaboration],
Phys.\ Rev.\ Lett.  {\bf 118}, 
251301 (2017)
[arXiv:1702.07666 [astro-ph.CO]].

\bibitem{Akerib:2017kat}
D.~S.~Akerib et al. [LUX Collaboration],
Phys.\ Rev.\ Lett.  {\bf 118}, 
251302 (2017)
[arXiv:1705.03380 [astro-ph.CO]].

\bibitem{Aprile:2017iyp} 
E.~Aprile {\it et al.} [XENON Collaboration],
Phys.\ Rev.\ Lett.\  {\bf 119}, 
181301 (2017)
[arXiv:1705.06655 [astro-ph.CO]].

\bibitem{Aprile:2018dbl} 
E.~Aprile {\it et al.} [XENON Collaboration],
Phys.\ Rev.\ Lett.\  {\bf 121}, no. 11, 111302 (2018)
[arXiv:1805.12562 [astro-ph.CO]].

\bibitem{Aartsen:2014oha}
M.~G.~Aartsen et al. [IceCube PINGU Collaboration],
arXiv:1401.2046 [physics.ins-det].

\bibitem{Choi:2015ara} 
K.~Choi et al. [Super-Kamiokande Collaboration],
Phys.\ Rev.\ Lett.\  {\bf 114}, 
141301 (2015)
[arXiv:1503.04858 [hep-ex]].

\bibitem{Aartsen:2016zhm}
M.~G.~Aartsen et al. [IceCube Collaboration],
Eur.\ Phys.\ J.\ C {\bf 77}, 
146 (2017)
[arXiv:1612.05949 [astro-ph.HE]].

\bibitem{Aguilar:2015ctt} 
M.~Aguilar et al. [AMS Collaboration],
211101 (2015).

\bibitem{TheFermi-LAT:2017vmf} 
M.~Ackermann et al. [Fermi-LAT Collaboration],
Astrophys.\ J.\  {\bf 840}, 
43 (2017)
[arXiv:1704.03910 [astro-ph.HE]].

\bibitem{Ambrosi:2017wek} 
G.~Ambrosi et al. [DAMPE Collaboration],
Nature {\bf 552}, 63 (2017)
[arXiv:1711.10981 [astro-ph.HE]].

\bibitem{Chen:2014oaa} 
C.~S.~Chen, F.~F.~Lee, G.~L.~Lin and Y.~H.~Lin,
JCAP {\bf 1410}, 049 (2014)
[arXiv:1408.5471 [hep-ph]].

\bibitem{Kong:2014mia} 
K.~Kong, G.~Mohlabeng and J.~C.~Park,
Phys.\ Lett.\ B {\bf 743}, 256 (2015)
[arXiv:1411.6632 [hep-ph]].

\bibitem{Chen:2015bwa} 
C.~S.~Chen, G.~L.~Lin and Y.~H.~Lin,
JCAP {\bf 1601}, 013 (2016)
[arXiv:1505.03781 [hep-ph]].

\bibitem{Chen:2015uha} 
J.~Chen, Z.~L.~Liang, Y.~L.~Wu and Y.~F.~Zhou,
JCAP {\bf 1512}, 021 (2015)
[arXiv:1505.04031 [hep-ph]].

\bibitem{Chen:2015poa} 
C.~S.~Chen, G.~L.~Lin and Y.~H.~Lin,
Phys.\ Dark Univ.\  {\bf 14}, 35 (2016)
[arXiv:1508.05263 [hep-ph]].

\bibitem{Catena:2016ckl} 
R.~Catena and A.~Widmark,
JCAP {\bf 1612}, 016 (2016)
[arXiv:1609.04825 [astro-ph.CO]].

\bibitem{Garani:2017jcj} 
R.~Garani and S.~Palomares-Ruiz,
JCAP {\bf 1705}, 007 (2017)
[arXiv:1702.02768 [hep-ph]].

\bibitem{Fornengo:2017lax} 
N.~Fornengo, A.~Masiero, F.~S.~Queiroz and C.~E.~Yaguna,
JCAP {\bf 1712}, 012 (2017)
[arXiv:1710.02155 [hep-ph]].

\bibitem{Chen:2018lsk} 
C.~S.~Chen and Y.~H.~Lin,
JHEP {\bf 1804}, 074 (2018)
[arXiv:1802.06956 [hep-ph]].

\bibitem{Gaidau:2018yws} 
C.~Gaidau and J.~Shelton,
JCAP {\bf 1906}, 022 (2019)
[arXiv:1811.00557 [hep-ph]].

\bibitem{Kouvaris:2007ay}
C.~Kouvaris,
Phys.\ Rev.\ D {\bf 77}, 023006 (2008)
[arXiv:0708.2362 [astro-ph]].

\bibitem{deLavallaz:2010wp}
A.~de Lavallaz and M.~Fairbairn,
Phys.\ Rev.\ D {\bf 81}, 123521 (2010)
[arXiv:1004.0629 [astro-ph.GA]].

\bibitem{Kouvaris:2010vv}
C.~Kouvaris and P.~Tinyakov,
Phys.\ Rev.\ D {\bf 82}, 063531 (2010)
[arXiv:1004.0586 [astro-ph.GA]].

\bibitem{Kouvaris:2010jy}
C.~Kouvaris and P.~Tinyakov,
Phys.\ Rev.\ D {\bf 83}, 083512 (2011)
[arXiv:1012.2039 [astro-ph.HE]].

\bibitem{Leung:2011zz} 
S.~C.~Leung, M.~C.~Chu and L.~M.~Lin,
Phys.\ Rev.\ D {\bf 84}, 107301 (2011)
[arXiv:1111.1787 [astro-ph.CO]].

\bibitem{McDermott:2011jp}
S.~D.~McDermott, H.~B.~Yu and K.~M.~Zurek,
Phys.\ Rev.\ D {\bf 85}, 023519 (2012)
[arXiv:1103.5472 [hep-ph]].

\bibitem{Kouvaris:2011gb}
C.~Kouvaris,
Phys.\ Rev.\ Lett.\  {\bf 108}, 191301 (2012)
[arXiv:1111.4364 [astro-ph.CO]].

\bibitem{Guver:2012ba}
T.~G\"{u}ver, A.~E.~Erkoca, M.~Hall Reno and I.~Sarcevic,
JCAP {\bf 1405}, 013 (2014)
[arXiv:1201.2400 [hep-ph]].

\bibitem{Bramante:2013nma}
J.~Bramante, K.~Fukushima, J.~Kumar and E.~Stopnitzky,
Phys.\ Rev.\ D {\bf 89}, 015010 (2014)
[arXiv:1310.3509 [hep-ph]].

\bibitem{Tolos:2015qra} 
L.~Tolos and J.~Schaffner-Bielich,
Phys.\ Rev.\ D {\bf 92}, 123002 (2015)
[arXiv:1507.08197 [astro-ph.HE]].

\bibitem{Bramante:2017xlb}
J.~Bramante, A.~Delgado and A.~Martin,
Phys.\ Rev.\ D {\bf 96}, 063002 (2017)
[arXiv:1703.04043 [hep-ph]].

\bibitem{Baryakhtar:2017dbj}
M.~Baryakhtar, J.~Bramante, S.~W.~Li, T.~Linden and N.~Raj,
Phys.\ Rev.\ Lett.\  {\bf 119},  131801 (2017)
[arXiv:1704.01577 [hep-ph]].

\bibitem{Raj:2017wrv}
N.~Raj, P.~Tanedo and H.~B.~Yu,
Phys.\ Rev.\ D {\bf 97}, 043006 (2018)
[arXiv:1707.09442 [hep-ph]].

\bibitem{Ellis:2017jgp} 
J.~Ellis, A.~Hektor, G.~H\"{u}tsi, K.~Kannike, L.~Marzola, M.~Raidal and V.~Vaskonen,
arXiv:1710.05540 [astro-ph.CO].

\bibitem{Ellis:2018bkr} 
J.~Ellis, G.~H\"{u}tsi, K.~Kannike, L.~Marzola, M.~Raidal and V.~Vaskonen,
arXiv:1804.01418 [astro-ph.CO].

\bibitem{Bell:2018pkk} 
N.~F.~Bell, G.~Busoni and S.~Robles,
JCAP {\bf 1809}, 018 (2018)
[arXiv:1807.02840 [hep-ph]].

\bibitem{Garani:2018kkd}
   R.~Garani, Y.~Genolini and T.~Hambye,
  JCAP {\bf 1905}, 035 (2019)
  [arXiv:1812.08773 [hep-ph]].

\bibitem{Hamaguchi:2019oev} 
K.~Hamaguchi, N.~Nagata and K.~Yanagi,
Phys.\ Lett.\ B {\bf 795}, 484 (2019)
[arXiv:1905.02991 [hep-ph]].

\bibitem{Dasgupta:2019juq} 
B.~Dasgupta, A.~Gupta and A.~Ray,
JCAP {\bf 1908}, 018 (2019)
[arXiv:1906.04204 [hep-ph]].

\bibitem{Acevedo:2019agu} 
J.~F.~Acevedo, J.~Bramante, R.~K.~Leane and N.~Raj,
arXiv:1911.06334 [hep-ph].

\bibitem{Joglekar:2019vzy} 
A.~Joglekar, N.~Raj, P.~Tanedo and H.~B.~Yu,
arXiv:1911.13293 [hep-ph].

\bibitem{Chen:2018ohx} 
C.~S.~Chen and Y.~H.~Lin,
JHEP {\bf 1808}, 069 (2018)
[arXiv:1804.03409 [hep-ph]].

\bibitem{Kaplan:2009ag} 
D.~E.~Kaplan, M.~A.~Luty and K.~M.~Zurek,
Phys.\ Rev.\ D {\bf 79}, 115016 (2009)
[arXiv:0901.4117 [hep-ph]].

\bibitem{Petraki:2013wwa} 
K.~Petraki and R.~R.~Volkas,
Int.\ J.\ Mod.\ Phys.\ A {\bf 28}, 1330028 (2013)
[arXiv:1305.4939 [hep-ph]].

\bibitem{Zurek:2013wia} 
K.~M.~Zurek,
Phys.\ Rept.\  {\bf 537}, 91 (2014)
[arXiv:1308.0338 [hep-ph]].

\bibitem{Kouvaris:2013kra} 
C.~Kouvaris and P.~Tinyakov,
Phys.\ Rev.\ D {\bf 90}, 043512 (2014)
[arXiv:1312.3764 [astro-ph.SR]].

\bibitem{Colpi:1986ye} 
M.~Colpi, S.~L.~Shapiro and I.~Wasserman,
Phys.\ Rev.\ Lett.\  {\bf 57}, 2485 (1986).

\bibitem{Boehmer:2007um} 
C.~G.~Boehmer and T.~Harko,
JCAP {\bf 0706}, 025 (2007)
[arXiv:0705.4158 [astro-ph]].

\bibitem{Kouvaris:2015rea} 
C.~Kouvaris and N.~G.~Nielsen,
Phys.\ Rev.\ D {\bf 92}, 063526 (2015)
[arXiv:1507.00959 [hep-ph]].

\bibitem{Eby:2015hsq} 
J.~Eby, C.~Kouvaris, N.~G.~Nielsen and L.~C.~R.~Wijewardhana,
JHEP {\bf 1602}, 028 (2016)
[arXiv:1511.04474 [hep-ph]].

\bibitem{Zheng:2014fya} 
H.~Zheng, K.~J.~Sun and L.~W.~Chen,
Astrophys.\ J.\  {\bf 800}, 
141 (2015)
[arXiv:1408.2926 [nucl-th]].

\bibitem{Tulin:2017ara}
S.~Tulin and H.~B.~Yu,
Phys. Rept. \textbf{730}, 1 (2018)
[arXiv:1705.02358 [hep-ph]].

\bibitem{Bullock:2017xww}
J.~S.~Bullock and M.~Boylan-Kolchin,
Ann. Rev. Astron. Astrophys. \textbf{55}, 343 (2017)
[arXiv:1707.04256 [astro-ph.CO]].

\bibitem{Randall:2007ph}
S.~W.~Randall, M.~Markevitch, D.~Clowe, A.~H.~Gonzalez and M.~Bradac,
Astrophys. J. \textbf{679}, 1173 (2008)
[arXiv:0704.0261 [astro-ph]].

\bibitem{Walker:2011zu}
M.~G.~Walker and J.~Penarrubia,
Astrophys. J. \textbf{742}, 20 (2011)
[arXiv:1108.2404 [astro-ph.CO]].

\bibitem{BoylanKolchin:2011de}
M.~Boylan-Kolchin, J.~S.~Bullock and M.~Kaplinghat,
Mon. Not. Roy. Astron. Soc. \textbf{415}, L40 (2011)
[arXiv:1103.0007 [astro-ph.CO]].

\bibitem{BoylanKolchin:2011dk}
M.~Boylan-Kolchin, J.~S.~Bullock and M.~Kaplinghat,
Mon. Not. Roy. Astron. Soc. \textbf{422}, 1203 (2012)
[arXiv:1111.2048 [astro-ph.CO]].

\bibitem{Elbert:2014bma}
O.~D.~Elbert, J.~S.~Bullock, S.~Garrison-Kimmel, M.~Rocha, J.~Oñorbe and A.~H.~Peter,
Mon. Not. Roy. Astron. Soc. \textbf{453}, 29 (2015)
[arXiv:1412.1477 [astro-ph.GA]].

\bibitem{yhl_git}
\texttt{https://github.com/yenhsunlin/dm2nsbh}

\bibitem{Buckley:2009in} 
M.~R.~Buckley and P.~J.~Fox,
Phys.\ Rev.\ D {\bf 81}, 083522 (2010)
[arXiv:0911.3898 [hep-ph]].

\bibitem{Tulin:2013teo} 
S.~Tulin, H.~B.~Yu and K.~M.~Zurek,
Phys.\ Rev.\ D {\bf 87}, 115007 (2013)
[arXiv:1302.3898 [hep-ph]].

\bibitem{Wise:2014jva} 
M.~B.~Wise and Y.~Zhang,
Phys.\ Rev.\ D {\bf 90}, 055030 (2014)
Erratum: [Phys.\ Rev.\ D {\bf 91}, 039907 (2015)]
[arXiv:1407.4121 [hep-ph]].

\bibitem{Kamada:2016euw}
A.~Kamada, M.~Kaplinghat, A.~B.~Pace and H.~B.~Yu,
Phys.\ Rev.\ Lett.\  {\bf 119}, 111102 (2017)
[arXiv:1611.02716 [astro-ph.GA]].

\bibitem{Robertson:2017mgj}
A.~Robertson {\it et al.},
arXiv:1711.09096 [astro-ph.CO].

\bibitem{Oman:2015xda}
K.~A.~Oman et al.,
Mon.\ Not.\ Roy.\ Astron.\ Soc.\  {\bf 452}, 
3650 (2015)
[arXiv:1504.01437 [astro-ph.GA]].

\bibitem{Elbert:2016dbb} 
O.~D.~Elbert, J.~S.~Bullock, M.~Kaplinghat, S.~Garrison-Kimmel, A.~S.~Graus and M.~Rocha,
Astrophys.\ J.\  {\bf 853}, 109 (2018)
[arXiv:1609.08626 [astro-ph.GA]].

\bibitem{Davoudiasl:2012ag} 
H.~Davoudiasl, H.~S.~Lee and W.~J.~Marciano,
Phys.\ Rev.\ D {\bf 85}, 115019 (2012)
[arXiv:1203.2947 [hep-ph]].
  
\bibitem{Kaplinghat:2013yxa} 
M.~Kaplinghat, S.~Tulin and H.~B.~Yu,
Phys.\ Rev.\ D {\bf 89}, 035009 (2014)
[arXiv:1310.7945 [hep-ph]].

\bibitem{Gould:1987ww} 
A.~Gould,
Astrophys.\ J.\  {\bf 328}, 919 (1988).

\bibitem{Gould:1987ir} 
A.~Gould,
Astrophys.\ J.\  {\bf 321}, 571 (1987).

\bibitem{Gould:1987ju} 
A.~Gould,
Astrophys.\ J.\  {\bf 321}, 560 (1987).

\bibitem{Busoni:2013kaa} 
G.~Busoni, A.~De Simone and W.~C.~Huang,
JCAP {\bf 1307}, 010 (2013)
[arXiv:1305.1817 [hep-ph]].

\bibitem{Bell:2013xk} 
N.~F.~Bell, A.~Melatos and K.~Petraki,
Phys.\ Rev.\ D {\bf 87}, 123507 (2013)
[arXiv:1301.6811 [hep-ph]].

\bibitem{priv} 
Private communication with the authors of Ref.~\cite{Garani:2018kkd}

\bibitem{Jungman:1995df} 
G.~Jungman, M.~Kamionkowski and K.~Griest,
Phys.\ Rept.\  {\bf 267}, 195 (1996)
[hep-ph/9506380].

\bibitem{Feng:2011vu} 
J.~L.~Feng, J.~Kumar, D.~Marfatia and D.~Sanford,
Phys.\ Lett.\ B {\bf 703}, 124 (2011)
[arXiv:1102.4331 [hep-ph]].

\bibitem{Lin:2014hla} 
G.~L.~Lin, Y.~H.~Lin and F.~F.~Lee,
Phys.\ Rev.\ D {\bf 91}, 033002 (2015)
[arXiv:1409.3094 [hep-ph]].

\bibitem{Bertoni:2013bsa} 
B.~Bertoni, A.~E.~Nelson and S.~Reddy,
Phys.\ Rev.\ D {\bf 88}, 123505 (2013)
[arXiv:1309.1721 [hep-ph]].

\bibitem{Gresham:2017zqi}
M.~I.~Gresham, H.~K.~Lou and K.~M.~Zurek,
Phys.\ Rev.\ D \textbf{96}, 096012 (2017)
[arXiv:1707.02313 [hep-ph]].

\bibitem{Gresham:2017cvl}
M.~I.~Gresham, H.~K.~Lou and K.~M.~Zurek,
Phys.\ Rev.\ D \textbf{97}, 036003 (2018)
[arXiv:1707.02316 [hep-ph]].

\bibitem{Gresham:2018anj}
M.~I.~Gresham, H.~K.~Lou and K.~M.~Zurek,
Phys.\ Rev.\ D \textbf{98}, 096001 (2018)
[arXiv:1805.04512 [hep-ph]].


\end{thebibliography}
\end{document}